\documentclass[pra,showpacs,aps,twocolumn,epsfig]{revtex4}
\usepackage{psfrag,graphicx}
\usepackage{amsmath,amssymb}
\newcommand{\ket}[1]{\left| #1 \right\rangle}
\newcommand{\bra}[1]{\left\langle #1 \right|}

\begin{document}
\title{Phase encoding schemes for measurement device independent quantum key distribution with basis-dependent flaw}
\author{Kiyoshi Tamaki $^{1, 2}$}
\author{Hoi-Kwong Lo$^{3}$}
\author{Chi-Hang Fred Fung$^{4}$}
\author{Bing Qi$^{3}$}

\affiliation{
$^{1}$NTT Basic Research Laboratories, NTT Corporation,\\
3-1,Morinosato Wakamiya Atsugi-Shi, Kanagawa, 243-0198, Japan\\
$^{2}$National Institute of Information and Communications Technology, 4-2-1 Nukui-Kita, Koganei, Tokyo 184-8795, Japan\\ 
$^{3}$Center for Quantum Information and Quantum Control,
Dept. of Electrical \& Computer Engineering and Dept. of Physics,
University of Toronto, Toronto, Ontario, M5S 3G4, Canada\\
$^{4}$Department of Physics and Center of Computational and Theoretical Physics,
University of Hong Kong, Pokfulam Road, Hong Kong\\
}
\date{\today}

\begin{abstract}
In this paper, we study the unconditional security of the so-called measurement device independent quantum key distribution (MDIQKD) with the basis-dependent flaw in the context of phase encoding schemes. We propose two schemes for the phase encoding, the first one employs a phase locking technique with the use of non-phase-randomized coherent pulses, and the second one uses conversion of standard BB84 phase encoding pulses into polarization modes. We prove the unconditional security of these schemes and we also simulate the key generation rate based on simple device models that accommodate imperfections. Our simulation results show the feasibility of these schemes with current technologies and highlight the importance of the state preparation with good fidelity between the density matrices in the two bases. Since the basis-dependent flaw is a problem not only for MDIQKD but also for standard QKD, our work highlights the importance of an accurate signal source in practical QKD systems.\\
{\bf Note: We include the erratum of this paper in Appendix C. The correction does not affect the validity of the main 
conclusions reported in the paper, which is the importance of the state preparation in MDIQKD and the fact that our 
schemes can generate the key with the practical channel mode that we have assumed.}
\end{abstract}
%\pacs{03.67.-a,89.70.-a}

\maketitle
\section{Introduction}
Quantum key distribution (QKD) is often said to be unconditionally secure \cite{M96, LC98, SP00}. More precisely, QKD can be proven to be secure against any eavesdropping {\it given} that the users' (Alice and Bob) devices satisfy some requirements, which often include mathematical characterization of users' devices as well as the assumption that there is no side-channel. This means that no one can break mathematical model of QKD, however in practice, it is very difficult for practical devices to meet the requirements, leading to the breakage of the security of practical QKD systems. Actually, some attacks on QKD have been proposed and demonstrated successfully against practical QKD systems \cite{hack, det}. 

To combat the practical attacks, some counter-measures \cite{theory-counter}, including device independent security proof idea \cite{device independent}, have been proposed. The device independent security proof is very interesting from the theoretical viewpoint, however it cannot apply to practical QKD systems where loopholes in testing Bell's inequality \cite{Bell} cannot be closed. As for the experimental counter-measures, battle-testing of the practical detection unit has attracted many researchers' attention \cite{det} since the most successful practical attack so far is to exploit the imperfections of the detectors.

Recently, a very simple and very promising idea, which is called a measurement device independent QKD (MDIQKD) has been proposed by Lo, Curty, and Qi \cite{LCQ11}. In this scheme, neither Alice nor Bob performs any measurement, but they only send out quantum signals to a measurement unit (MU). MU is a willing participant of the protocol, and MU can be a network administrator or a relay. However, MU can be untrusted and completely under the control of the eavesdropper (Eve). After Alice and Bob send out signals, they wait for MU's announcement of whether she has obtained the successful detection, and proceed to the standard post-processing of their sifted data, such as error rate estimation, error correction, and privacy amplification. The basic idea of MDIQKD is based on a reversed EPR-based QKD protocol \cite{reversed}, which is equivalent to EPR-based QKD \cite{EPR} in the sense of the security, and MDIQKD is remarkable because it removes {\it all} the potential loopholes of the detectors without sacrificing the performance of standard QKD since Alice and Bob do not detect any quantum signals from Eve. Moreover, it is shown in \cite{LCQ11} that MDIQKD with infinite number of decoy states and polarization encoding can cover about twice the distance of standard decoyed QKD, which is comparable to EPR-based QKD. The only assumption needed in MDIQKD is that the preparation of the quantum signal sources by Alice and Bob is (almost) perfect and carefully characterized. We remark that the characterization of the signal source should be easier than that of the detection unit since the characterization of the detection unit involves the estimation of the response of the devices to unknown input signals sent from Eve.

%This idea is based on the so-called reversed-EDP (Entanglement-Distillation-Protocol), and the equivalence among the reversed EDP, the standard EDP, and the actual protocol ensures the security of the actual protocol since a pure state, from which key is distilled, is virtually distilled.

With MDIQKD in our hand, we do not need to worry about imperfections of MU any more, and we should focus our attention more to the imperfections of signal sources. One of the important imperfections of the sources is the basis-dependent flaw that stems from the discrepancy of the density matrices corresponding to the two bases in BB84 states. The security of standard BB84 with basis-dependent flaw has been analyzed in \cite{GLLP, Koashi05, LP06} which show that the basis-dependent flaw decreases the achievable distance. Thus, in order to investigate the practicality of MDIQKD, we need to generalize the above works to investigate the security of MDIQKD under the imperfection. Another problem in MDIQKD is that the first proposal is based on polarization encoding \cite{LCQ11}, however, in some situations where birefringence effect in optical fiber is highly time-dependent, we need to consider MDIQKD with phase encoding rather than polarization encoding. In this paper, we study the above issues simultaneously.

We first propose two schemes of the phase encoding MDIQKD, one employs phase locking of two separate laser sources and the other one uses the conversion of phase encoding into polarization encoding. Then, we prove the unconditional security of these schemes with basis-dependent flaw by generalizing the quantum coin idea \cite{GLLP, LP06, Koashi05}. Based on the security proof, we simulate the key generation rate with realistic parameters, especially we employ a simple model to evaluate the basis-dependent flaw due to the imperfection of the phase modulators. Our simulation results imply that the first scheme covers shorter distances and may require less accuracy of the state preparation, while the second scheme can cover much longer distances when we can prepare the state very precisely. We note that in this paper we consider the most general type of attacks allowed
by quantum mechanics and establish unconditional security for our protocols.

This paper is organized as follows. In Sec. \ref{sec:protocol}, we give a generic description of MDIQKD protocol, and we propose our schemes in Sec. \ref{sec:Phase encoding scheme I} and Sec. \ref{sec:Phase encoding scheme II}. Then, we prove the unconditional security of our schemes in Sec. \ref{sec:proof}, and we present some simulation results of the key generation rate based on realistic parameters in Sec. \ref{sec:simulation}. Finally, we summarize this paper in Sec. \ref{sec:summary}.

\section{Protocol}\label{sec:protocol}
In this section, we introduce MDIQKD protocol whose description is generic for all the schemes that we will introduce in the following sections.
The MDIQKD protocol runs as follows.

Step (1): Each of Alice and Bob prepares a signal pulse and a reference pulse, and each of Alice and Bob applies phase modulation to the signal pulse, which is randomly chosen from $0$, $\pi/2$, $\pi$, and $3\pi/2$. Here, $\{0, \pi\}$ ($\{\pi/2, 3\pi/2\}$) defines  $X$ ($Y$)-basis. Alice and Bob send both pulses through quantum channels to Eve who possesses MU.

Step (2): MU performs some measurement, and announces whether the measurement outcome is successful or not. It also broadcasts whether the successful event is the detection of type-0 or type-1 (The two types of the successful outcomes correspond to two specific Bell states \cite{Bell state}).

Step (3): If the measurement outcome is successful, then Alice and Bob keep their data. Otherwise, they discard the data. When the outcome is successful, Alice and Bob broadcast their bases and they keep the data only when the bases match, which we call sifted key. Depending on the type of the successful event and the basis that they used, Bob may or may not perform bit-flip on his sifted key.

Step (4): Alice and Bob repeat (1)-(3) many times until they have large enough number of the sifted key.

Step (5): They sacrifice a portion of the data as the test bits to estimate the bit error rate and the phase error rate on the remaining data (code bits).

Step (6): If the estimated bit error and phase error rates are too high, then they abort the protocol, otherwise they proceed.

Step (7): Alice and Bob agree over a public channel on an error correcting code and on a hash function depending on the bit and phase error rate on the code bits. After performing error correction and privacy amplification, they share the key.

The role of the MU in Eve is to establish a quantum correlation, i.e., a Bell state, between Alice and Bob to generate the key. If it can establish the strong correlation, then Alice and Bob can generate the key, and if it cannot, then it only results in a high bit error rate to be detected by Alice and Bob and they abort the protocol. As we will see later, since Alice and Bob can judge whether they can generate a key or not by only checking the experimental data as well as information on the fidelity between the density matrices in $X$- basis and $Y$-basis, it does not matter who performs the measurement nor what kind of measurement is actually done as long as MU broadcasts whether the measurement outcome was successful together with the information of whether the successful outcome is type-0 or type-1. 

In the security proof, we assume that MU is totally under the control of Eve. In practice, however, we should choose an appropriate measurement that establishes the strong correlation under the normal operation, i.e., the situation without Eve who induces the channel losses and noises. In the following sections, we will propose two phase encoding MDIQKD schemes.

\section{Phase encoding scheme I}\label{sec:Phase encoding scheme I}

\begin{figure}
\begin{center}
 \includegraphics[scale=0.45]{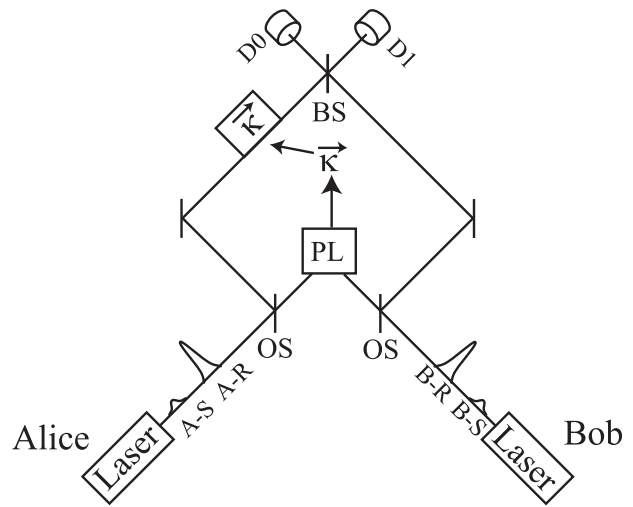}
 \end{center}
 \caption{Schematics of an experimental setup for the phase encoding scheme I. A-S (B-S) and A-R (B-R) respectively represents Alice's (Bob's) signal and reference pulses. The signal pulses are phase modulated according to Alice's and Bob's choice. OS represents an optical switch, which allows the reference pulse and the signal pulse to be transmitted and to be reflected, respectively. PL represents an unit measuring relative phase of two orthogonal polarization modes and it outputs the two relative phase information $\vec{\kappa}$. Then, the phase shift of $\vec{\kappa}$ for each polarization mode is applied to one of the signal pulses, and they will be detected by D0 and D1 after the interference at the 50:50 beam splitter BS. \label{setup1}}
\end{figure}

In this section, we propose an experimental setup for MDIQKD with phase encoding scheme, which is depicted in Fig. \ref{setup1}. This scheme will be proven to be unconditionally secure, i.e., secure against the most general type of attacks allowed by quantum mechanics. In this setup, we assume that the intensity of Alice's signal (reference) pulse matches with that of Bob's signal (reference) pulse when they enter MU. In order to lock the relative phase, we use strong pulses as the reference pulses. In PL unit in the figure, the relative phase between the two strong pulses is measured in two polarization modes separately. The measurement result is denoted by $\vec{\kappa}$ (here, the arrow represents two entries that correspond to the two relative phases). Depending on this information $\vec{\kappa}$, appropriate phase modulations for two polarization modes are applied to incoming signal pulse from Alice. Then, Alice's and Bob's signal pulses are input into the 50/50 beam splitter which is followed by two single-photon threshold detectors. The successful event of type-0 (type-1) in step (2) is defined as the event where only D0 (D1) clicks. In the case of type-1 successful detection event, Bob applies bit flip to his sifted key (we define the phase relationship of BS in such a way that D1 never clicks when the phases of the two input signal coherent pulses are the same). 

Roughly speaking, our scheme performs double BB84 \cite{BB84}, i.e., each of Alice and Bob is sending signals in the BB84 states, without phase randomization \cite{LP06}. Differences between our scheme and the polarization encoding MDIQKD scheme include that Alice and Bob do not need to share the reference frame for the polarization mode, since MU performs the feed-forward control of the polarization, and our scheme intrinsically possesses the basis-dependent flaw.

To see how this particular setup establishes the quantum correlation under the normal operation, it is convenient to consider an entanglement distribution scheme \cite{BDSW}, which is mathematically equivalent to the actual protocol. For the simplicity of the discussion, we assume the perfect phase locking for the moment and we only consider the case where both of Alice and Bob use $X$-basis. We skip the discussion for $Y$-basis, however it holds in a similar manner \cite{Y basis}. In this case, the actual protocol is equivalently described as follows. First, Alice prepares two systems in the following state, which is a purification of the $X$-basis density matrix, 
\begin{eqnarray}
\ket{\phi_{x}^{(+)} (\ket{\sqrt{\alpha_{A}}})}&\equiv&\frac{1}{\sqrt{2}}\Big(\ket{0_x}_{A1}\ket{\sqrt{\alpha}_{A}}_{A2}\nonumber\\
&+&\ket{1_x}_{A1}\ket{-\sqrt{\alpha_{A}}}_{A2}\Big)
\label{def-state}
\end{eqnarray}
and sends the second system to MU through the quantum channel. Here, $\ket{\sqrt{\alpha_{A}}}_{A2}$ and $\ket{-\sqrt{\alpha_{A}}}_{A2}$ represent coherent states that Alice prepares in the actual protocol ($\alpha_{A}$ represents the mean photon number or inetensity), $\ket{0_x}$ and $\ket{1_x}$ are eigenstate of the computational basis ($X$ basis), which is related with $Y$-basis eigenstate through $\ket{0_y}\equiv (i\ket{0_x}+\ket{1_x})/\sqrt{2}$ and $\ket{1_y}\equiv (\ket{0_x}+i\ket{1_x})/\sqrt{2}$. For the later convenience, we also define $Z$-basis states as $\ket{0_z}\equiv(\ket{0_x}+\ket{1_x})/\sqrt{2}$ and $\ket{1_z}\equiv(\ket{0_x}-\ket{1_x})/\sqrt{2}$. Moreover, the subscript of $x$ in $\ket{\phi_{x}^{(+)} (\ket{\sqrt{\alpha_{A}}})}$ represents that Alice is to measure her qubit along $X$-basis, the subscript of $A$ in $\alpha_{A}$ refers to the party who prepares the system, and the superscript $(+)$ represents the relative phase of the superposition. Similarly, Bob also prepares two systems in a similar state $\ket{\phi_{x}^{(+)} (\ket{\sqrt{\alpha_{B}}})}$, sends the second system to MU, and performs $X$-basis measurement. Note that $X$-basis measurement by Alice and Bob can be delayed after Eve's announcement of the successful event without losing any generalities in the security analysis, and we assume this delay in what follows.

In order to see the joint state of the qubit pair after the announcement, note that the beam splitter converts the joint state $\ket{\phi_{x}^{(+)} (\ket{\sqrt{\alpha_{A}}})}\ket{\phi_{x}^{(+)} (\ket{\sqrt{\alpha_{B}}})}$ into the following state $\ket{\zeta}_{A1, B1, {\rm D0}, {\rm D1}}$ 
\begin{eqnarray}
\ket{\zeta}_{A1, B1, {\rm D0}, {\rm D1}}&\equiv&\frac{1}{2}\Big(\ket{0_x}_{A1}\ket{0_x}_{B1}\ket{\sqrt{2\alpha'}}_{{\rm D0}}\ket{{0}}_{{\rm D1}}\nonumber\\
&+&\ket{1_x}_{A1}\ket{1_x}_{B1}\ket{-\sqrt{2\alpha'}}_{{\rm D0}}\ket{{0}}_{{\rm D1}}\nonumber\\
&+&\ket{0_x}_{A1}\ket{1_x}_{B1}\ket{{0}}_{{\rm D0}}\ket{\sqrt{2\alpha'}}_{{\rm D1}}\nonumber\\
&+&\ket{1_x}_{A1}\ket{0_x}_{B1}\ket{{0}}_{{\rm D0}}\ket{-\sqrt{2\alpha'}}_{{\rm D1}}\Big)\,.\nonumber\\
\label{normal schI}
\end{eqnarray}
Here, for the simplicity of the discussion, we assume that there is no channel losses, we define $\alpha_{A}=\alpha_{B}\equiv\alpha'$, and $\ket{{0}}$ represents the vacuum state. Moreover, the subscripts ${\rm D0}$ and ${\rm D1}$ represent the output ports of the beam splitter. If detector D0 (D1) detects photons and the other detector D1 (D0) detects the vacuum state, i.e., type-0 (type-1) event, it is shown in the Appendix A that the joint probability of having type-0 (type-1) successful event and Alice and Bob share the maximally entangled state $\ket{\Psi^{+}}$ ($\ket{\Psi^{-}}$) is $(1-e^{-4 \alpha'})/4$. We note that since $\ket{\sqrt{2\alpha'}}\neq\ket{-\sqrt{2\alpha'}}$, Alice and Bob do not always share this state, and with a joint probability of $(1-e^{-2 \alpha'})^2/4$, they have type-0 (type-1) successful event and share the maximally entangled state with the phase error, i.e., the bit error in $Y$-basis, as $\ket{\Phi^{+}}$ ($\ket{\Phi^{-}}$).

Note that the bit-flip operation in type-1 successful detection can be equivalently performed by $\pi$ rotation around $Z$-basis before Bob performs $X$ basis measurement. In other words, $\pi$ rotation around $Z$-basis before $X$-basis measurement does not change the statistics of the $X$-basis measurement followed by the bit-flip. Thanks to this property, we can conclude that Alice and Bob share $\ket{\Psi^{+}}$ with probability of $(1-e^{-4 \alpha'})/2$ and $\ket{\Phi^{+}}$ with probability of $(1-e^{-2 \alpha'})^2/2$ after the rotation. This means that even if Alice and Bob are given the successful detection event, they cannot be sure whether they share $\ket{\Phi^{+}}$ or $\ket{\Psi^{+}}$, however, if they choose a small enough $\alpha$, then the phase error rate (the rate of the state $\ket{\Phi^{+}}$ in the qubit pairs remaining after the successful events or equivalently, the rate of $Y$-basis bit error among all the shared qubit pairs) becomes small and they can generate a pure state $\ket{\Psi^{+}}$ by phase error correction, which is equivalently done by privacy amplification in the actual protocol \cite{SP00}. We note that the above discussion is valid only for the case without noises and losses, and we will prove the security against the most general attack in Sec. \ref{sec:proof} without relying on the argument given in this section.

We remark that in the phase encoding scheme I, it is important that Alice and Bob know quite well about the four states that they prepare. This may be accomplished by using state tomography with homodyne measurement involving the use of the strong reference pulse \cite{homodyne}.

\section{Phase encoding scheme II}\label{sec:Phase encoding scheme II}
\label{scheme II}

\begin{figure*}
\begin{center}
 \includegraphics[scale=0.55]{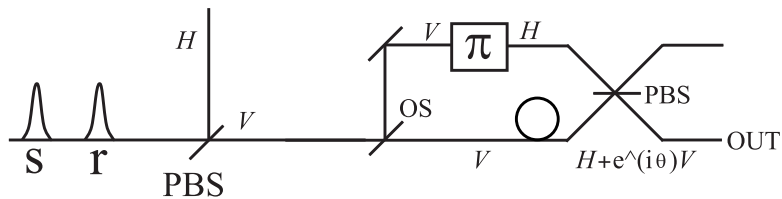}
 \end{center}
 \caption{Schematics of an experimental setup of the converter from phase encoding to polarization encoding. PBS is a polarization beam splitter, OS represents an optical switch that routes the reference pulse and signal pulses to different paths. The ``$\pi$'' performs the conversion $\ket{H}\rightarrow\ket{V}$. The circle represents time-delay. The italic characters along the lines represent the polarization state.  \label{setup2}}
\end{figure*}

\begin{figure*}
\begin{center}
 \includegraphics[scale=0.55]{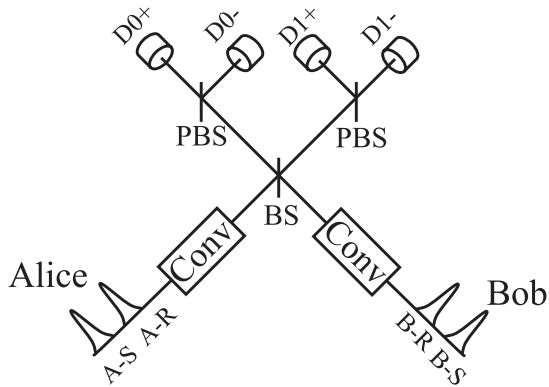}
 \end{center}
 \caption{Schematics of an experimental setup of MU. A-S (B-S) and A-R (B-R) respectively represent Alice's (Bob's) signal and reference pulses, and MU consists of two converters for each pulse from Alice and Bob (depicted as ``Conv''), and Bell measurement unit consists of a 50:50 beam splitter (BS) followed by two polarization beam splitters (PBSs). See the main text for the explanation. \label{Bell M}}
\end{figure*}

In this section, we propose the second experimental setup for MDIQKD with phase encoding scheme. Like scheme I, this scheme will also be proven to be unconditionally secure. In this scheme, the coherent pulses that Alice and Bob send out are exactly the same as those in the standard phase encoding BB84, i.e., $ \ket{e^{i(\zeta+\theta)}\sqrt{\alpha}}_{s}\ket{e^{i\zeta}\sqrt{\alpha}}_{r}$ where subscripts $s$ and $r$ respectively denote the signal pulse and the reference pulse, $\zeta$ is a completely random phase, $\theta$ is randomly chosen from $\{0, \pi/2, \pi, 3\pi/2\}$ to encode the information. After entering the MU, each pulse pair is converted from a phase coding signal to a polarization coding signal by a phase-to-polarization converter (see details below). We note that thanks to the phase randomization by $\zeta$, the joint state of the signal pulse and the reference pulse is a classical mixture of photon number states.

In Fig. \ref{setup2}, we show the schematics of the converter. This converter performs the phase-to-polarization conversion: ${\hat P}_{1} \ket{e^{i(\zeta+\theta)}\sqrt{\alpha}}_{s}\ket{e^{i\zeta}\sqrt{\alpha}}_{r}$ to $(\ket{V}+e^{i\theta}\ket{H})/\sqrt{2}$, where ${\hat P}_{1}$ is a projector that projects the joint system of the signal and reference pulses to a two-dimensional single-photon subspace spanned by $\{\ket{0}_{s}\ket{1}_{r}, \ket{1}_{s}\ket{0}_{r}\}$ where $0$ and $1$ represent the photon number, and $\ket{H}$ ($\ket{V}$) represents the horizontal (vertical) polarization state of a single-photon. To see how it works, let us follow the time evolution of the input state. At the polarization beam splitter (PBS in Fig. \ref{setup2}), the signal and reference pulses first split into two polarization modes, H and V, and we throw away the pulses being routed to V mode. Then, in H mode, the signal pulse and the reference pulse are routed to different paths by using an optical switch, and we apply $\pi$-rotation only to one of the paths to convert H to V. At this point, we essentially have $(\ket{V}_{\rm up}+e^{i\theta}\ket{H}_{\rm lw})/\sqrt{2}$, where the subscripts of ``${\rm up}$'' and ``${\rm lw}$'' respectively denote the upper path and the lower path. Finally, these spatial modes ${\rm up}$ and ${\rm lw}$ are combined together by using a polarization beam splitter so that we have $(\ket{V}+e^{i\theta}\ket{H})/\sqrt{2}$ in the output port depicted as ``OUT''.

In practice, since the birefringence of the quantum channel can be highly time dependent and the polarization state of the input pulses to MU may randomly change with time, i.e., the input polarization state is a completely mixed state, we cannot deterministically distill a pure polarization state, and thus the conversion efficiency can never be perfect. In other words, one may consider the same conversion of the V mode just after the first polarization beam splitter, however it is impossible to combine the resulting polarization pulses from V mode and the one from H mode into a single mode. 

We assume that MU has two converters, one is for the conversion of Alice's pulse and the other one is for Bob's pulse, and the two output ports ``OUT'' are connected to exactly the same Bell measurement unit \cite{Bell} in the polarization encoding MDIQKD scheme in Fig. \ref{Bell M} \cite{LCQ11}. This Bell measurement unit consists of a 50:50 beam splitter, two polarization beam splitters, and four single-photon detectors, which only distinguishes perfectly two out of the four Bell states of $\ket{\Phi^{-}}$ and $\ket{\Psi^{-}}$. The polarization beam splitters discriminate between $\ket{+}\equiv(\ket{H}+\ket{V})/\sqrt{2}$ and $\ket{-}\equiv(\ket{H}-\ket{V})/\sqrt{2}$ (note that we choose $+$ and $-$ modes rather than H and V modes since our computational basis is $+$ and $-$). Suppose that a single-photon enters both from Alice and Bob. In this case, the click of D0+ and D0- or D1+ and D1- means the detection of $\ket{\Phi^{-}}$, and the click of D0+ and D1- or D0- and D1+ means the detection of $\ket{\Psi^{-}}$ (see Fig. \ref{Bell M}). In this scheme, since the use of coherent light induces non-zero bit error rate in $Y$-basis ($\{(\ket{H}+i\ket{V})/\sqrt{2}, (\ket{H}-i\ket{V})/\sqrt{2}\}$-basis), we consider to generate the key from $\{\ket{+}, \ket{-}\}$-basis and we use the data in $Y$-basis only to estimate the bit error rate in this basis conditioned on that both of Alice and Bob emit a single-photon, which determines the amount of privacy amplification. By considering a single-photon polarization input both from Alice and Bob, one can see that Bob should not apply the bit flip only when Alice and Bob use $Y$-basis and $\Phi^{-}$ is detected in MU, and Bob should apply the bit flip in all the other successful events to share the same bit value. Accordingly, the bit error in $X$-basis is given by the successful detection event conditioned on that Alice and Bob's polarization are identical. As for $Y$-basis, the bit error is $\Phi^{-}$ detection given the orthogonal polarizations or $\Psi^{-}$ detection given the identical polarization.

Assuming completely random input polarization state, our converter successfully converts the single-photon pulse with a probability of $50\%$. Note in the normal experiment that the birefringence effect between Alice and the converter and the one between Bob and the converter are random and independent, however it only leads to fluctuating coincidence rate of Alice's and Bob's signals at the Bell measurement, but does not affect the QBER. Moreover, the fluctuation increases the single-photon loss inserted into the Bell measurement. Especially, the events that the output of the converter for Alice is the vacuum and the one for Bob is a single-photon, and vice versa would increase compared to the case where we have no birefringence effect. However, this is not a problem since the Bell measurement does not output the conclusive events in these cases unless the dark counting occurs. Thus, the random and independent polarization fluctuation in the normal experiment is not a problem, and we will simply assume in our simulation in Sec. \ref{simulationII} that this fluctuation can be modeled just by $50\%$ loss. We emphasize that we do not rely on these assumptions at all when we prove the security, and our security proof applies to any channels and MUs.

For the better performance and also for the simplicity of analysis, we assume the use of infinite number of decoy states \cite{decoy} to estimate the fraction of the probability of successful event conditioned on that both of Alice and Bob emit a single-photon. One of the differences in our analysis from the work in \cite{LCQ11} is that we will take into account the imperfection of Alice's and Bob's source, i.e., the decay of the fidelity between two density matrices in two bases. We also remark that since the H and V modes are defined locally in MU, Alice and Bob do not need to share the reference frame for the polarization mode, which is one of the qualitative differences from polarization encoding MIQKD scheme \cite{LCQ11}.

\section{Security proof}\label{sec:proof}
This section is devoted to the unconditional security proof, i.e., the security proof against the most general attacks, of our schemes. Since both of our schemes are based on BB84 and the basis-dependent flaw in both protocols can be treated in the same manner, we can prove the security in a unified manner.

If the states sent by Alice and Bob were basis independent, i.e., the density matrices of $X$-basis and $Y$-basis were the same, then the security proof of the original BB84 \cite{M96, LC98, SP00} could directly apply (also see \cite{intuition} for a bit more detailed discussion of this proof), however they are basis dependent in our case. Fortunately, security proof of standard BB84 with basis-dependent flaw has already been shown to be secure \cite{GLLP, Koashi05, LP06}, and we generalize this idea to our case where we have basis-dependent flaw from both of Alice and Bob. In order to do so, we consider a virtual protocol \cite{GLLP, Koashi05, LP06, virtual} that Alice and Bob get together and the basis choices by Alice and Bob are made via measurement processes on the so-called quantum coin. In this virtual protocol of the phase encoding scheme I, Alice and Bob prepare joint systems in the state \cite{Y basis2}
\begin{eqnarray}
&&\ket{\Psi'}\nonumber\\
&\equiv&\frac{1}{2}\Big(\ket{0_z}_{E}\ket{0_z}_{B}\ket{0_z}_{A}\ket{\phi_{x}^{(+)} (\ket{\sqrt{\alpha_{A}}})}\ket{\phi_{x}^{(+)} (\ket{\sqrt{\alpha_{B}}})}\nonumber\\
&+&\ket{0_z}_{E}\ket{0_z}_{B}\ket{1_z}_{A}\ket{\phi_{y}^{(+)} (\ket{-i\sqrt{\alpha_{A}}})}\ket{\phi_{y}^{(+)} (\ket{-i\sqrt{\alpha_{B}}})}\nonumber\\
&+&\ket{1_z}_{E}\ket{1_z}_{B}\ket{0_z}_{A}\ket{\phi_{x}^{(+)} (\ket{\sqrt{\alpha_{A}}})}\ket{\phi_{y}^{(+)} (\ket{-i\sqrt{\alpha_{B}}})}\nonumber\\
&+&\ket{1_z}_{E}\ket{1_z}_{B}\ket{1_z}_{A}\ket{\phi_{y}^{(+)} (\ket{-i\sqrt{\alpha_{A}}})}\ket{\phi_{x}^{(+)} (\ket{\sqrt{\alpha_{B}}})}\Big)\,.\nonumber\\
\label{coin-joint-state}
\end{eqnarray}
Since just replacing the state, for instance $\ket{\phi_{x}^{(+)} (\ket{\sqrt{\alpha_{A}}})}\rightarrow\ket{\phi_{x}^{(+)} (\ket{1}_{s}\ket{0}_{r}/\sqrt{2}+\ket{0}_{s}\ket{1}_{r}/\sqrt{2})}$ where $1$ and $0$ in the ket respectively represents the single-photon and the vacuum, is enough to apply the following proof to the phase encoding scheme II, we discuss only the security of the phase encoding scheme I in what follows. In Eq. (\ref{coin-joint-state}), the first system denoted by $E$ is given to Eve just after the preparation, and it informs Eve of whether the bases to be used by Alice and Bob match or not. The second system, denoted by $B$, is a copy of the first system and this system is given to Bob who measures this system with $\{\ket{0_z}_{B}, \ket{1_z}_{B}\}$ basis to know whether Alice's and Bob's bases match or not. If his measurement 
outcome is $|0_{z}\rangle_{B}$ ($|1_{z}\rangle_{B}$), then he uses the same 
(the other) basis to be used by Alice (note that no classical communication is needed 
in order for Bob to know Alice's basis since Alice and Bob get together). The third system, which is denoted by $A$ and we call `` quantum coin'', is possessed and to be measured by Alice along $\{\ket{0_z}_{A}, \ket{1_z}_{A}\}$ basis to determine her basis choice, and the measurement outcome will be 
sent to Eve after Eve broadcasts the measurement outcome at MU. Moreover, all the second systems of $\ket{\phi_{x}^{(+)} (\sqrt{\alpha_{A}})}$, $\ket{\phi_{y}^{(+)} (\ket{-i\sqrt{\alpha_{A}}})}$, $\ket{\phi_{x}^{(+)} (\ket{\sqrt{\alpha_{B}}})}$, and $\ket{\phi_{y}^{(+)} (\ket{-i\sqrt{\alpha_{B}}})}$ are sent to Eve. Note in this formalism that the information, including classical information and quantum information, available to Eve is the same as those in the actual protocol, and the generated key is also the same as the one of the actual protocol since the statistics of Alice's and Bob's raw data is exactly the same as the one of the actual protocol. Thus, we are allowed to work on this virtual protocol for the security proof. 

The first system given to Eve in Eq. (\ref{coin-joint-state}) allows her to know which coherent pulses contain data in the sifted key and she can post-select only the relevant pulses. Thus, without the loss of any generalities of the security proof, we can concentrate only on the post-selected version of the state in Eq. (\ref{coin-joint-state}) as 
\begin{eqnarray}
& &\ket{\Psi}\equiv\frac{1}{\sqrt{2}}\Big(\ket{0_z}_{A}\ket{\phi_{x}^{(+)} (\ket{\sqrt{\alpha_{A}}})}\ket{\phi_{x}^{(+)} (\ket{\sqrt{\alpha_{B}}})}\nonumber\\
&+&\ket{1_z}_{A}\ket{\phi_{y}^{(+)} (\ket{-i\sqrt{\alpha_{A}}})}\ket{\phi_{y}^{(+)} (\ket{-i\sqrt{\alpha_{B}}})}\Big)\,.
\label{coin-joint-state-post-selected}
\end{eqnarray}

The most important quantity in the proof is the phase error rate in the code bits. The definition of the phase error rate is the rate of bit errors along $Y$-basis in the sifted key if they had chosen $Y$-basis as the measurement basis when both of them have sent pulses in $X$-basis. If Alice and Bob have a good estimation of this rate as well as the bit error rate in the sifted key (the bit error rate in $X$-basis given Alice and Bob have chosen $X$-basis for the state preparation), they can perform hashing in $Y$-basis and $X$-basis simultaneously \cite{BDSW, TK10} to distill pairs of qubits in the state whose fidelity with respect to the product state of the maximally entangled state $\ket{\Psi^{+}}$ is close to $1$.  

According to the discussion on the universal composability \cite{composable}, the key distilled via $X$-basis measurement on such a state is composably secure and moreover exactly the same key can be generated only by classical means, i.e., error correction and privacy amplification \cite{SP00}. Thus, we are left only with the phase error estimation. For the simplicity of the discussion, we assume the large number of successful events $n$ so that we neglect all the statistical fluctuations and we are allowed to work on a probability rather than the relative frequency.

The quantity we have to estimate is the bit error along $Y$-basis, denoted by $\delta_{y}'$, given Alice and Bob send $\ket{\phi_{x}^{(+)}(\ket{\sqrt{\alpha_{A}}})}\ket{\phi_{x}^{(+)} (\ket{\sqrt{\alpha_{B}}})}$ state, which is different from the experimentally available bit error rate along $Y$-basis given Alice and Bob send $\ket{\phi_{y}^{(+)}(\ket{-i\sqrt{\alpha_{A}}})}\ket{\phi_{y}^{(+)} (\ket{-i\sqrt{\alpha_{B}}})}$ state. Intuitively, if the basis-dependent flaw is very small, $\delta_{y}'$ and $\delta_{y}$ should be very close since the states are almost indistinguishable. To make this intuition rigorous, we briefly review the idea by \cite{Koashi05, LP06} which applies Bloch sphere bound \cite{TKI03} to the quantum coin. Suppose that we randomly choose $Z$-basis or $X$-basis as the measurement basis for each quantum coin. Let $n\gamma_{z}/2$ and $n\gamma_{x}/2$ be fraction that those quantum coins result in $1$ in $Z$-basis and $X$-basis measurement, respectively. What Bloch sphere bound, i.e., Eq. (13) or Eq. (14) in \cite{TKI03} or Eq. (A1) in \cite{LP06}, tells us in our case is that no matter how the correlations among the quantum coins are and no matter what the state for the quantum coins is, thanks to the randomly chosen bases, the following inequality holds with probability exponentially close to $1$ in $n$,
\begin{eqnarray}
(1-2\gamma_{z})^2+(1-2\gamma_{x})^2\le1\,.
\end{eqnarray}
By applying this bound separately to the quantum coins that are conditional on having phase errors and to those that are conditional on having no phase error, and furthermore by combining those inequalities using Bayes's rule, we have
\begin{eqnarray}
1-2\Delta\le\sqrt{\delta_{y}\delta_{y}'}+\sqrt{(1-\delta_{y})(1-\delta_{y}')}\,.
\label{phase error bound}
\end{eqnarray}
Here, $\Delta$ is equivalent to the probability that the measurement outcome of the quantum coin along $X$-basis is $\ket{1_x}$ given the successful event in MU. Note that this probability can be enhanced by Eve who chooses carefully the pulses, and Eve could attribute all the loss events to the quantum coins being in the state $\ket{0_x}$. Thus, we have an upper bound of $\Delta$ in the worst case scenario as
\begin{eqnarray}
\Delta\le\Delta_{\rm ini}/\gamma_{\rm suc}\,,
\label{fDelta}
\end{eqnarray}
and 
\begin{eqnarray}
\Delta_{\rm ini}&\equiv& \Big(1-\Big\langle\phi_{x}^{(+)} (\ket{\sqrt{\alpha_{A}}}) \Big|\phi_{y}^{(+)} (\ket{-i\sqrt{\alpha_{A}}})\Big\rangle\nonumber\\
&\times& \Big\langle\phi_{x}^{(+)} (\ket{\sqrt{\alpha_{B}}})\Big|\phi_{y}^{(+)} (\ket{-i\sqrt{\alpha_{B}}})\Big\rangle\Big)/2\,,
\label{delta}
\end{eqnarray}
where $\gamma_{\rm suc}$ is the frequency of the successful event.

Note that we have not used the explicit form of $|\phi_{x}^{(\pm)} (\beta)\rangle$ and $|\phi_{y}^{(\pm)} (\beta)\rangle$, where $\beta=\sqrt{\alpha_{A}}, - i\sqrt{\alpha_{A}}, \sqrt{\alpha_{B}}, -i\sqrt{\alpha_{B}}$, in the derivation of Eqs. (\ref{phase error bound}), (\ref{fDelta}), and (\ref{delta}), and the important point is that the state $|\phi_{x}^{(\pm)} (\beta)\rangle$ and $|\phi_{y}^{(\pm)} (\beta)\rangle$ are the purification of Alice's and Bob's density matrices for both bases. Since there always exists purification states of $\rho^{(X)}$ and $\rho^{(Y)}$, which are respectively denoted by $\ket{\Omega^{(X)}}$ and $\ket{\Omega^{(Y)}}$, such that $\left\langle\Omega^{(X)}|\Omega^{(Y)}\right\rangle=F(\rho^{(X)}, \rho^{(Y)})\equiv {\rm Tr}\left(\left|\sqrt{\rho^{(X)}}\sqrt{\rho^{(Y)}}\right|\right)$, $\Delta_{\rm ini}$ can be rewritten by
\begin{eqnarray}
\Delta_{\rm ini}&\equiv& \Big[1-F\left(\rho_{A}^{(X)}, \rho_{A}^{(Y)}\right)F\left(\rho_{B}^{(X)}, \rho_{B}^{(Y)}\right)\Big]/2\,,
\label{fidDelta}
\end{eqnarray}
where $\rho_{A}^{(X)}$ represents Alice's density matrix of $X$ basis and all the other density matrices are defined by the same manner. Our expression of $\Delta_{\rm ini}$ has the product of two fidelities, while the standard BB84 with basis-dependent flaw in \cite{GLLP, Koashi05, LP06} has only one fidelity (the fidelity between Alice's density matrices in $X$ and $Y$ bases). The two products may lead to poor performance of our schemes compared to that of standard QKD in terms of the achievable distances, however our schemes have the huge advantage over the standard QKD that there is no side-channel in the detectors.

Finally, the key generation rate $G$, given $X$-basis, in the asymptotic limit of large $n$ is given by
\begin{eqnarray}
G=\gamma_{\rm suc}^{(x)}\left(1-f(\delta_{x})h(\delta_{x})-h(\delta_{y}')\right)\,,
\label{key rate}
\end{eqnarray}
where $\delta_{x}$ is the bit error rate in $X$-basis, $f(\delta_{x})$ is the inefficiency of the error correcting code, and $h(x)\equiv -x\log_{2} x-(1-x)\log_{2} (1-x)$. We can trivially obtain the key generation rate for $Y$-basis just by interchanging $X$-basis in all the discussions above to $Y$-basis. We remark in our security proof that we have assumed nothing about what kind of measurement MU conducts but that it announces whether it detects the successful event and the type of the event (this announcement allows us to calculate $\gamma_{\rm suc}^{(x)}$ and the error rates). Thus, MU can be assumed to be totally under the control of Eve.

%%%%%%%%%%%%%%%%%
\section{Simulation of the key generation rates}\label{sec:simulation}
%%%%%%%%%%%%%%%%%

In the following subsections, we show some examples of the key generation rate of each of our schemes assuming typical experimental parameters taken from Gobby-Yuan-Shields (GYS) experiment \cite{GYS} unless otherwise stated. Moreover, we assume that the imperfect phase modulation is the main source of the decay of the fidelity between the density matrices in two bases, and we evaluate the effect of this imperfection on the key generation rate.
 
\subsection{Phase encoding scheme I}
In the phase encoding scheme I, the important quantity for the security $\Delta_{\rm ini}$ can be expressed as
\begin{eqnarray}
\Delta_{\rm ini}&=&\frac{1}{2}\Big[1-e^{-(\alpha_{A}+\alpha_{B})}(\cos\alpha_{A}+\sin\alpha_{A})\nonumber\\
&\times&(\cos\alpha_{B}+\sin\alpha_{B})\Big]\,.
\label{delta11}
\end{eqnarray}
Note that this quantity is dependent on the intensity of Alice's and Bob's sources. As we have mentioned in Sec. III, this quantity may be estimated relatively easily via tomography involving homodyne measurement. 

\begin{figure}
\begin{center}
 \includegraphics[scale=0.6]{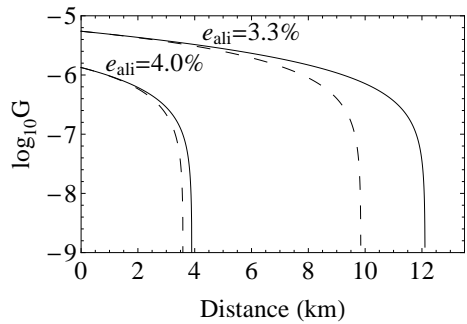}
 \end{center}
 \caption{The key generation rates of each setting as a function of the distance between Alice and Bob with the alignment error rate ($e_{\rm ali}$) of $3.3\%$ and $4.0\%$. Dashed line: (a) MU is at Bob's side, i.e., $l_{B}=0$. Solid line: (b) MU is just in the middle between Alice and Bob. The lines achieving the longer distances correspond to $3.3\%$ of $e_{\rm ali}$. See also the main text for the explanation. \label{fig:key33}}
\end{figure}

\begin{figure}
\begin{center}
 \includegraphics[scale=0.6]{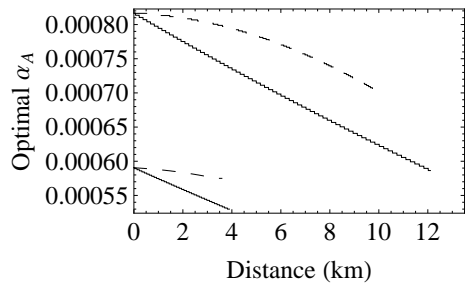}
 \end{center}
 \caption{Optimal mean photon numbers emitted by Alice's source ($\alpha_{A}$) that outputs Fig. \ref{fig:key33} as a function of the distance between Alice and Bob. \label{fig:intensity33}}
\end{figure}

To simulate the resulting key generation rate, we assume that the bit error stems from the dark counting as well as alignment errors due to imperfect phase locking or imperfect optical components. The alignment error is assumed to be proportional to the probability of having a correct click caused only by the optical detection not by the dark counting. Moreover, we make assumptions that all the detectors have the same characteristics for the simplicity of the analysis, and Alice and Bob choose the intensities of the signal lights in such a way that the intensities of the incoming pulses to MU are the same. Finally, we assume the quantum inefficiency of the detectors to be part of the losses in the quantum channels. With all the assumptions, we may express the resulting experimental parameters as
\begin{eqnarray}
\gamma_{\rm suc}^{(x)}&=&[p_{\rm dark}+(1-p_{\rm dark})(1-e^{-2\alpha_{\rm in}})](1-p_{\rm dark})\nonumber\\
&+&(1-p_{\rm dark})e^{-2\alpha_{\rm in}}p_{\rm dark}\nonumber\\
\gamma_{\rm suc}&=&\gamma_{\rm suc}^{(x)}+\gamma_{\rm suc}^{(y)}\nonumber\\
\delta_x&=&\delta_y=\Big[e_{\rm ali}(1-p_{\rm dark})^2(1-e^{-2\alpha_{\rm in}})\nonumber\\
&+&(1-p_{\rm dark})e^{-2\alpha_{\rm in}}p_{\rm dark}\Big]/\gamma_{\rm suc}^{(x)}\nonumber\\
\alpha_{\rm in}&\equiv&\alpha_{A}\eta_{A}=\alpha_{B}\eta_{B}\nonumber\\
\eta_{A}&=&\eta_{{\rm det}, A}10^{-\xi_{A} l_{A}/10}\nonumber\\
\eta_{B}&=&\eta_{{\rm det}, B}10^{-\xi_{B} l_{B}/10}\,.
\label{ex data I}
\end{eqnarray} 
Here, $p_{\rm dark}$ is the dark count rate of the detector, $e_{\rm ali}$ is the alignment error rate, $\eta_{A} (\eta_{B})$ is Alice's (Bob's) overall transmission rate, $\eta_{{\rm det}, A}$ ($\eta_{{\rm det}, B}$) is the quantum efficiency of Alice's (Bob's) detector, $\xi_{A} (\xi_{B})$ is Alice's (Bob's) channel transmission rate, and $l_{A}$ ($l_{B}$) is the distance between Alice (Bob) and MU. The first term and the second term in $\delta_x$ or $\delta_y$ respectively represent the alignment error, which is assumed to be proportional to the probability of having correct bit value due to the detection of the light, and errors due to dark counting (one detector clicks due to the dark counting while the other one does not). 

We take the following parameters from GYS experiment \cite{GYS}: $f(\delta_{x})=1.22$, $p_{\rm dark}=8.5\times10^{-7}$, $\xi=0.21$ (dB/km), $\eta_{{\rm det}, A}=\eta_{{\rm det}, B}=0.045$, and $e_{\rm ali}=0.033$, and we simulate the key generation rate as a function of the distance between Alice and Bob in Fig. \ref{fig:key33}. In the figure, we consider two settings: (a) MU is at Bob's side, i.e., $l_{B}=0$ (b) MU is just in the middle between Alice and Bob. The reason why we consider these setting is that the basis-dependent flaw is dependent on intensities that Alice and Bob employ, and it is not trivial where we should place MU for the better performance. %Note in (c) that we replace $\alpha_{B}$ in Eq. (\ref{delta11}) with $\alpha_{\rm in}$, and thus the basis-dependent flaw from Bob's side is almost negligible as $\alpha_{\rm in}\ll\alpha_{B}$. 

Since MDIQKD polarization encoding scheme without basis-dependent flaw achieves almost twice the distance of BB84 \cite{LCQ11}, we may expect that the setting (b) could achieve almost twice the distance of BB84 without phase randomization that achieves about 13 (km) \cite{LP06} with the same experimental parameters. The simulation result, however, does not follow this intuition since we have the basis-dependent flaw not only from Alice's side but also from Bob's side. Thus, the advantage that we obtain from putting MU between Alice and Bob is overwhelmed by the double basis-dependent flaw. In each setting, we have optimized the intensity of the coherent pulses $\alpha_{A}$ for each distance (see Fig. \ref{fig:intensity33}). 

In order to explain why the optimal $\alpha_A$ is so small, note that scheme I intrinsically suffers from the basis-dependent flaw due to Eq. (\ref{delta11}). This means that if we use relatively large $\alpha_A$, then we cannot generate the key due to the flaw. Actually, when we set $\alpha_A=0.1$, which is a typical order of the amplitude for decoy BB84, one can see that the upper bound of the phase error rate is $1/2$ even in the zero distance, i.e., $l=0$, and we have no chance to generate the key with this amplitude. Thus, Alice and Bob have to reduce the intensities in order to suppress the basis-dependent flaw. Also, as the distance gets larger and the losses get increased, Alice and Bob have to use weaker pulses since larger losses can be exploited by Eve to enhance the basis-dependent flaw according to Eq. (\ref{fDelta}), and they can reduce the intensities until it reaches the cut-off value where the detection of the weak pulses is overwhelmed by the dark counts.

In the above simulation, we have assumed that Alice and Bob can prepare states very accurately, however in reality, they can only prepare approximate states due to the imperfection of the sources. This imperfection gives more basis-dependent flaw, and in order to estimate the effect of this imperfection, we assume that the fidelity between the two actually prepared density matrices in two bases is approximated by the fidelity between the following density matrices (see Appendix B for the detail)
\begin{eqnarray}
\rho_{X}^{(\rm Act)}(\alpha, \delta)=\left(\ket{\sqrt{\alpha}}\bra{\sqrt{\alpha}}+\ket{-e^{i|\delta|}\sqrt{\alpha}}\bra{-e^{i|\delta|}\sqrt{\alpha}}\right)/2\nonumber\\
\end{eqnarray}
and 
\begin{eqnarray}
\rho_{Y}^{(\rm Act)}(\alpha, \delta)&=&\Big(\ket{i e^{i|\delta|/2}\sqrt{\alpha}}\bra{i e^{i|\delta|/2}\sqrt{\alpha}}\nonumber\\
&+&\ket{-i e^{-i|\delta|/2}\sqrt{\alpha}}\bra{-i e^{-i|\delta|/2}\sqrt{\alpha}}\Big)/2\,,\nonumber\\
\end{eqnarray}
where we assume an imperfect phase modulator whose degree of the phase modulation error is proportional to the target phase modulation value, and $\delta$ represents the imperfection of the phase modulation that is related with the extinction ratio $\eta_{\rm ex}$ as  
\begin{eqnarray}
\left|\tan\frac{\delta}{2}\right|^2=\eta_{\rm ex}\,.
\label{mimperfect phase modulator}
\end{eqnarray}
In this equation, we assume that the non-zero extinction ratio is only due to the imperfection of the phase modulators. Since imperfect phase modulation results in the same effect as the alignment errors, i.e., the pulses are routed to a wrong output port, we assume that the alignment error rate is increased with this imperfection. Thus, in the simulation accommodating the imperfection of the phase modulation, we replace $e_{\rm ali}$ with $e_{\rm ali}+16\eta_{\rm ex}$. Here, we have used a pessimistic assumption that the effect of the phase modulation becomes $16$-times higher than before since each of Alice and Bob has one phase modulator and MU has two phase modulators for the phase shift of two polarization modes (note from Eq. (\ref{mimperfect phase modulator}) that $\eta_{\rm ex}$ is approximately proportional to $\delta^2$, thus 4 times degradation in terms of the accuracy of the phase modulation results in $16$-times degradation in terms of the extinction ratio). We also remark that in practice, it is more likely that the phase encoding errors are independent, in which case a factor of 4 will suffice and the key rate will actually be higher than what is presented in our paper. On the other hand, we have to use the following $\Delta_{\rm ini}$ when we consider the security: 
\begin{eqnarray}
\Delta_{\rm ini}&=&\Big[1-F(\rho_{X}^{(\rm Act)}(\alpha_{A}, \delta), \rho_{Y}^{(\rm Act)}(\alpha_{A}, \delta))\nonumber\\
&\times&F(\rho_{X}^{(\rm Act)}(\alpha_{B}, \delta), \rho_{Y}^{(\rm Act)}(\alpha_{B}, \delta))\Big]/2\,.
\end{eqnarray}

\begin{figure}
\begin{center}
 \includegraphics[scale=0.6]{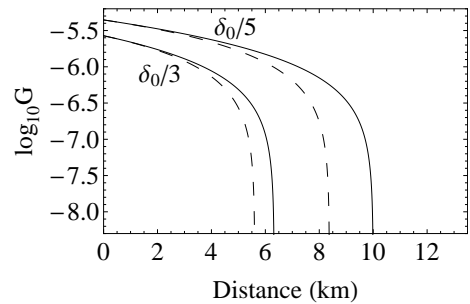}
 \end{center}
 \caption{The key generation rates of the each setting as a function of the distance between Alice and Bob with a baseline alignment error rate ($e_{\rm ali}$) of $3.3\%$ and imperfect phase modulators. $\delta_0=0.063$ represents the typical amount of the phase modulation error, and we plot the key rate for smaller imperfection of $\delta_{0}/3$ and $\delta_{0}/5$. Dashed line: MU is at Bob's side, i.e., $l_{B}=0$. Solid line: MU is just in the middle between Alice and Bob.\label{fig:keyimpferfectPMI}} 
\end{figure}

\begin{figure}
\begin{center}
 \includegraphics[scale=0.6]{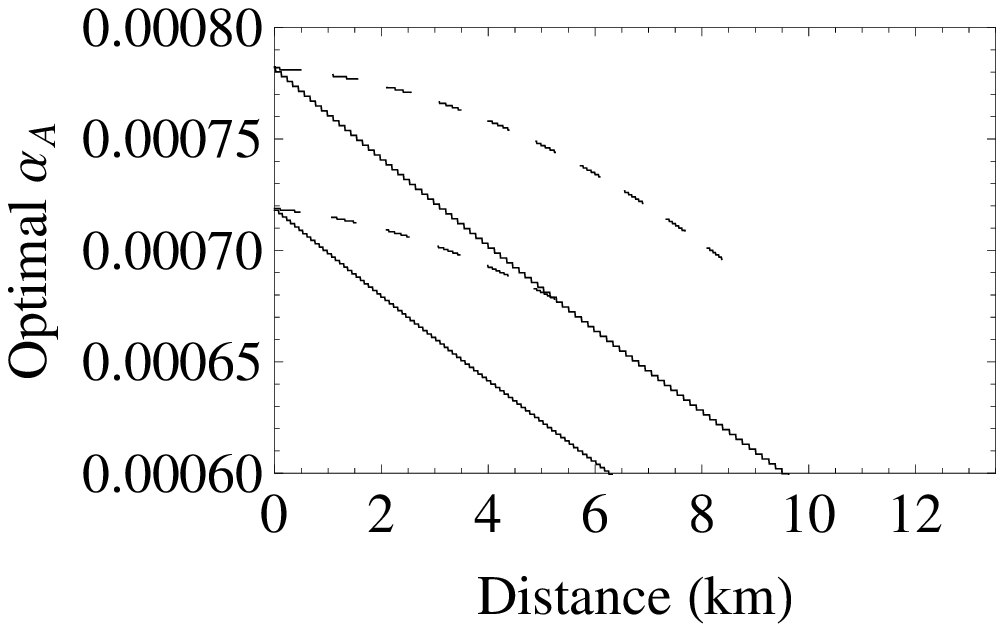}
 \end{center}
 \caption{Optimal mean photon numbers emitted by Alice's source ($\alpha_{A}$) that outputs Fig. \ref{fig:keyimpferfectPMI} as a function of the distance between Alice and Bob. \label{fig:intensityimpferfectPMI}}
\end{figure}

In Figs. \ref{fig:keyimpferfectPMI} and \ref{fig:intensityimpferfectPMI}, we plot the key generation rate and the corresponding optimal Alice's mean photon numbers ($\alpha_{A}$) as a function of the distance between Alice and Bob. In the figures, we define $|\delta|$ that satisfies $\eta_{\rm ex}=\left|\tan\frac{\delta}{2}\right|^2=10^{-3}$ as $\delta_0(\sim0.063)$, where $\eta_{\rm ex}=10^{-3}$ is the typical order of $\eta_{\rm ex}$ in some experiments \cite{extinction}. We have confirmed that we cannot generate the key when $\eta_{\rm ex}=10^{-3}$. However, we can see in the figures that if the accuracy of the phase modulation is increased three times or five times, i.e., $\delta=\delta_{0}/3$ and $\delta=\delta_{0}/5$, then we can generate the key. Like the case in Fig. \ref{fig:intensity33}, the small optimal mean photon number can be intuitively understood by the arguments that we have already made in this section.

In order to investigate the feasibility of the phase encoding scheme I with the current technologies, we replace $p_{\rm dark}=8.5\times10^{-7}$, $\eta_{{\rm det}, A}=\eta_{{\rm det}, B}=0.045$, and $e_{\rm ali}=0.033$ with $p_{\rm dark}=1.0\times10^{-7}$, $\eta_{{\rm det}, A}=\eta_{{\rm det}, B}=0.15$ \cite{uqcc}, and $e_{\rm ali}=0.0075$ \cite{LCQ11}. We see in Fig. \ref{fig:keyimpferfectPMInew} that the key generation is possible over much longer distances with those parameters assuming the precise control of the intensities of the laser source. We also show the corresponding optimal mean photon number $\alpha_A$ in Fig. \ref{fig:intensityimpferfectPMInew}. We note that thanks to the higher quantum efficiency, the success probability becomes higher, following that Alice and Bob can use larger mean photon number $\alpha_A$ compared to those in Figs. \ref{fig:intensityimpferfectPMI} and \ref{fig:intensityimpferfectPMInew}.

\begin{figure}
\begin{center}
 \includegraphics[scale=0.6]{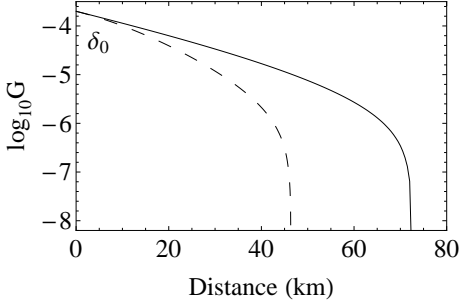}
 \end{center}
 \caption{The key generation rates of the each setting as a function of the distance between Alice and Bob with the latest parameters such as $e_{\rm ali}=0.0075$ with $p_{\rm dark}=1.0\times10^{-7}$, $\eta_{{\rm det}, A}=\eta_{{\rm det}, B}=0.15$ \cite{uqcc}, and $\delta_0=0.063$. Dashed line: MU is at Bob's side, i.e., $l_{B}=0$. Solid line: MU is just in the middle between Alice and Bob.\label{fig:keyimpferfectPMInew}}
\end{figure}

\begin{figure}
\begin{center}
 \includegraphics[scale=0.6]{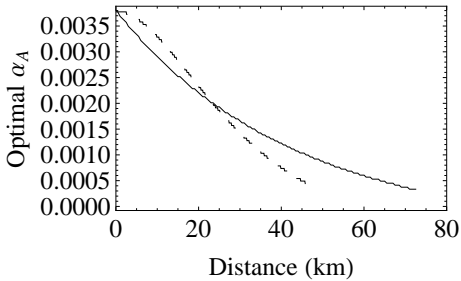}
 \end{center}
 \caption{Optimal mean photon numbers emitted by Alice's source ($\alpha_{A}$) that outputs Fig. \ref{fig:keyimpferfectPMInew} as a function of the distance between Alice and Bob. \label{fig:intensityimpferfectPMInew}}
\end{figure}

\subsection{Phase encoding scheme II}
\label{simulationII}
In the phase encoding scheme II, note that we can generate the key only from the successful detection event in MU given both of Alice and Bob send out a single-photon since if either or both of Alice and Bob emit more than one photon, then Eve can employ the so-called photon number splitting attack \cite{PNS}. Thus, the important quantities to estimate are $Q^{(1,1)}_{x}$, $\delta_{y}^{(1,1)}$, $\delta_{x}$, $Q_{x}$, which respectively represents gain in $X$-basis given both of Alice and Bob emit a single-photon, the phase error rate given Alice and Bob emit a single-photon, overall bit error rate in $X$-basis, and overall gain in $X$-basis. To estimate these quantities stemming from the simultaneous single-photon emission, we assume the use of infinite number of decoy states for the simplicity of analysis \cite{decoy}. Another important quantity in our study is the fidelity $F_{A}^{(1)}$ ($F_{B}^{(1)}$) between Alice's (Bob's) $X$-basis and $Y$-basis density matrices of only single-photon component, {\it not} whole optical modes. If this fidelity is given, then we have
\begin{eqnarray}
\Delta_{\rm ini}^{(1,1)}&=&\frac{1}{2}\left(1-F_{A}^{(1)}F_{B}^{(1)}\right)\,.
\end{eqnarray}
For the simplicity of the discussion, we consider the case of $F_{A}^{(1)}=F_{B}^{(1)}\equiv F^{(1)}$ in our simulation. The estimation of the fidelity only in the single-photon part is very important, however to the best of our knowledge we do not know any experiment directly measuring this quantity. This measurement may require photon number resolving detectors and very accurate interferometers. Thus, we again assume that the degradation of the fidelity is only due to the imperfect phase modulation given by Eq. (\ref{mimperfect phase modulator}), and we presume that the fidelity of the two density matrices between the two bases is approximated by the fidelity between the following density matrices  (see Appendix B for the detail)
\begin{eqnarray}
\rho_{X}^{(1)}&=&\frac{1}{2}\left[{\hat P}\left(\frac{\ket{0_z}+\ket{1_z}}{\sqrt{2}}\right)+{\hat P}\left(\frac{\ket{0_z}-e^{i|\delta|}\ket{1_z}}{\sqrt{2}}\right)\right]\nonumber\\
\rho_{Y}^{(1)}&=&\frac{1}{2}\Big[{\hat P}\left(\frac{\ket{0_z}+i e^{i|\delta|/2}\ket{1_z}}{\sqrt{2}}\right)\nonumber\\
&+&{\hat P}\left(\frac{\ket{0_z}-i e^{-i|\delta|/2}\ket{1_z}}{\sqrt{2}}\right)\Big]\,.
\end{eqnarray}
With these parameters, we can express the key generation rate given Alice and Bob use $X$-basis as \cite{GLLP}
\begin{eqnarray}
G=Q^{(1,1)}_{x}\left[1-h(\delta_{y}^{(1,1)'})\right]-f(\delta_{x})Q_{x}h(\delta_{x})\,,
\end{eqnarray}
where $\delta_{y}^{(1,1)'}$ is the $(1,1)$ version of $\delta_{y}'$ in Eq. (\ref{key rate}).

To simulate the resulting key generation rate, the bit errors are assumed to stem from multi-photon component, the dark counting, and the misalignment that is assumed to be proportional to the probability of obtaining the correct bit values only due to the detection by optical pulses. Like before, we also assume that all the detectors have the same characteristics, Alice and Bob choose the intensities of the signal lights in such a way that the intensities of the incoming pulses to MU are the same, and all the quantum inefficiencies of the detectors can be attributed to part of the losses in the quantum channel. Finally, Alice's and Bob's coherent light sources are assumed to be phase randomized, and the imperfect phase modulation is represented by the increase of the alignment error rate. With these assumptions, we may have the following resulting experimental parameters

\begin{eqnarray}
Q^{(1,1)}_{x}&=&4\alpha_{A}\alpha_{B}\eta_{A}\eta_{B}e^{-2(\alpha_{A}+\alpha_{B})}\nonumber\\
&\times&\Big[\frac{(1-p_{\rm dark})^2}{2}+\frac{p_{\rm dark}(1-p_{\rm dark})^2}{2}\Big]\nonumber\\
&+&W^{(2,1)}+W^{(2,0)}\nonumber\\
\delta_{x}^{(1,1)}&=&\Big\{4\alpha_{A}\alpha_{B}\eta_{A}\eta_{B}e^{-2(\alpha_{A}+\alpha_{B})}p_{\rm dark}(1-p_{\rm dark})^2/2\nonumber\\
&+&2(e_{\rm ali}+4\eta_{\rm ex})\alpha_{A}\alpha_{B}\eta_{A}\eta_{B}e^{-2(\alpha_{A}+\alpha_{B})}(1-p_{\rm dark})^2\nonumber\\
&+&(W^{(2,1)}+W^{(2,0)})/2\Big\}/Q^{(1,1)}_{x}\nonumber\\
Q^{(1,1)}_{y}&=&Q^{(1,1)}_{x}\nonumber\\
\delta^{(1,1)}_{y}&=&\delta^{(1,1)}_{x}\nonumber\\
W^{(2,1)}&\equiv&8\alpha_{A}\alpha_{B}e^{-2(\alpha_{A}+\alpha_{B})}\Big[\eta_{A}(1-\eta_{B})+(1-\eta_{A})\eta_{B}\Big]\nonumber\\
&\times&p_{\rm dark}(1-p_{\rm dark})^2\nonumber\\
W^{(2,0)}&\equiv&16\alpha_{A}\alpha_{B}(1-\eta_{A})(1-\eta_{B})e^{-2(\alpha_{A}+\alpha_{B})}\nonumber\\
&\times&p_{\rm dark}^2(1-p_{\rm dark})^2\nonumber\\
Q_{x}&=&2\left[1-(1-p_{\rm dark})e^{-\alpha_{\rm in}}\right]^2(1-p_{\rm dark})^2e^{-2\alpha_{\rm in}}+V \nonumber\\
\delta_{x}&=&V+(e_{\rm ali}+4\eta_{\rm ex})2\left(1-e^{-\alpha_{\rm in}}\right)^2 \nonumber\\
&\times&(1-p_{\rm dark})^2e^{-2\alpha_{\rm in}}\nonumber\\
V&\equiv&\frac{p_{\rm dark}(1-p_{\rm dark})}{2\pi}\nonumber\\
&\times&\int_{0}^{2\pi}d\theta\Big[1-(1-p_{\rm dark})e^{-\alpha_{\rm in}|1+e^{i\theta}|^2}\Big] \nonumber\\
&\times&\Big[(1-p_{\rm dark})e^{-\alpha_{\rm in}|1-e^{i\theta}|^2}\Big]\nonumber\\
&+&\frac{p_{\rm dark}(1-p_{\rm dark})}{2\pi}\nonumber\\
&\times&\int_{0}^{2\pi}d\theta\Big[1-(1-p_{\rm dark})e^{-\alpha_{\rm in}|1-e^{i\theta}|^2}\Big] \nonumber\\
&\times&\Big[(1-p_{\rm dark})e^{-\alpha_{\rm in}|1+e^{i\theta}|^2}\Big]\nonumber\\
\alpha_{\rm in}&\equiv&\alpha_{A}\eta_{A}=\alpha_{B}\eta_{B}\nonumber\\
\eta_{A}&=&\eta_{{\rm det}, A}10^{-\xi_{A} l_{A}/10}/2\nonumber\\
\eta_{B}&=&\eta_{{\rm det}, B}10^{-\xi_{B} l_{B}/10}/2\,
\label{exp-data-II}
\end{eqnarray} 
Note that $\alpha_{A}$ ($\alpha_{B}$) represents each of the intensity of Alice's (Bob's) signal light and the reference light, {\it not} the total intensity of them, and $\eta_{A}$ and $\eta_{B}$ are divided by $2$ since the conversion efficiency of our converter is $50\%$. $4$ in $4\eta_{\rm ex}$ again comes from the pessimistic assumption that each of Alice's and Bob's phase modulator is imperfect, and $W^{(2,1)}$ ($W^{(2,0)}$) represents the probability of the event where both of Alice and Bob emit a single-photon and only one (zero) photon is detected but the successful detection event is obtained due to the dark counting. On the other hand, the quantity that quantifies the basis-dependent flaw $\Delta$ in the present case is upper bounded by
\begin{eqnarray} 
\Delta&\le&\Delta_{\rm ini}^{(1,1)}/\left[Q^{(1,1)}/(4\alpha_{A}\alpha_{B}e^{-2(\alpha_{A}+\alpha_{B})})\right]\nonumber\\
Q^{(1,1)}&\equiv&(Q^{(1,1)}_{x}+Q^{(1,1)}_{y})/2
\end{eqnarray} 
where $Q^{(1,1)}/4\alpha_{A}\alpha_{B}e^{-2(\alpha_{A}+\alpha_{B})}$ is the probability that MU receives a single-photon both from Alice and Bob simultaneously conditioned on that each of Alice and Bob sends out a single-photon. We remark that $\Delta_{\rm ini}$ in this scheme is only dependent on the accuracy of the phase modulation. This is different from scheme I where the manipulation of the intensities of the pulses can affect the basis-dependent flaw.

In the simulation, we again assume GYS experimental parameters and we consider two settings: (a) MU is at Bob's side and (b) MU is just in the middle between Alice and Bob. Note that $\Delta_{\rm ini}$ is independent of $\alpha_{A}$ and $\alpha_{B}$ in the phase encoding scheme II case. 

\begin{figure}
\begin{center}
 \includegraphics[scale=0.6]{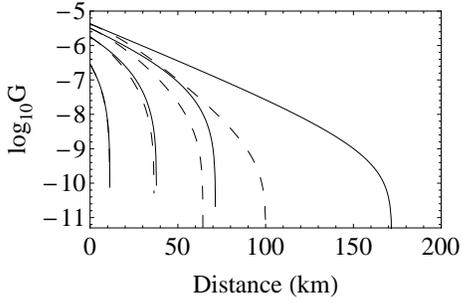}
 \end{center}
 \caption{The key generation rates of each setting as a function of the distance. Dashed line: (a) MU is at Bob's side, i.e., $l_{B}=0$. Solid line: (b) MU is just in the middle between Alice and Bob. We plot the key generation rates of each case when $\delta=0$, $\delta=\delta_{0}/50$, $\delta=\delta_{0}/20$, $\delta=\delta_{0}/10$ where $\delta$ is proportional to the amount of the phase modulation error, and for each case of (a) and (b) the key generation rates monotonously increase with the decrease of $\delta$., i.e., with the improvement of the phase modulation. The key rates of (a) and (b) when $\delta_{0}/10$ are almost superposed. See also the main text for the explanation. \label{F1}}
\end{figure}

In Fig. \ref{F1}, we plot the key generation rates of (a) and (b) for $\delta=0$, $\delta=\delta_{0}/50$, $\delta=\delta_{0}/20$, $\delta=\delta_{0}/10$ (recall from Eq. (\ref{mimperfect phase modulator}) that $\delta_{0}\sim0.0063224$ that corresponds to the typical extinction ratio of $0.1\%$), which respectively correspond to $F^{(1)}=1.0$, $F^{(1)}\sim1-1.0\times10^{-7}$, $F^{(1)}\sim 1-6.6\times10^{-7}$, and $F^{(1)}\sim 1-2.5\times10^{-6}$, and the achievable distances of (a) and (b) increase with the improvement of the accuracy, i.e., with the decrease of $\delta$. We have confirmed that no key can be distilled in (a) and (b) when $\delta\ge\delta_{0}/7$. The figure shows that the achievable distance drops significantly with the degradation of the accuracy of the phase modulator, and the main reason of this fast decay is that $\Delta$ is approximated by $\Delta_{\rm ini}/O(\eta_{A}\eta_{B})$ and this dominator decreases exponentially with the increase of the distance. 

\begin{figure}
\begin{center}
 \includegraphics[scale=0.57]{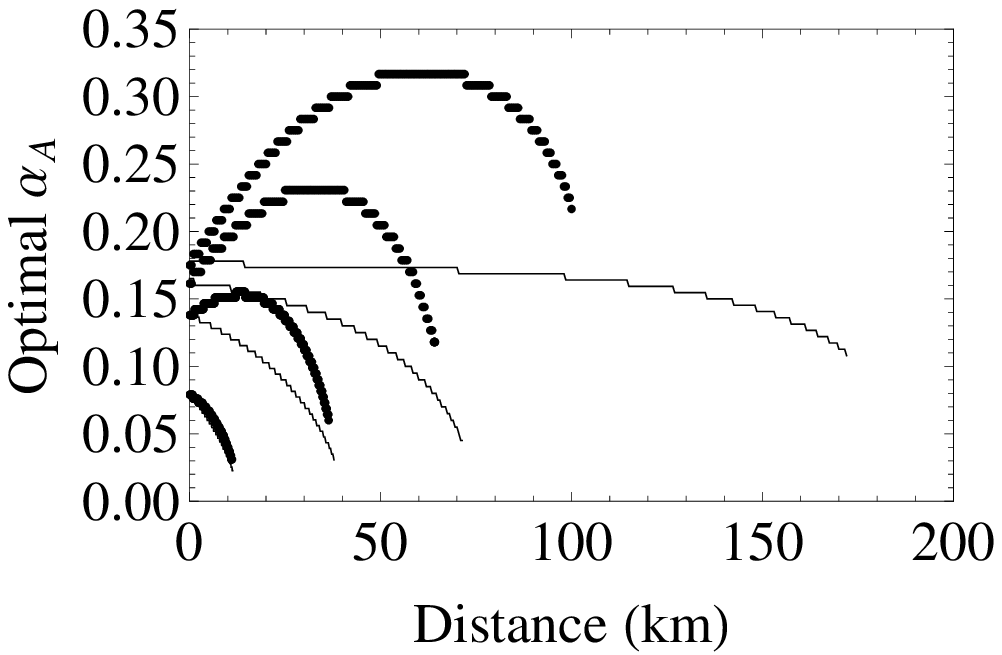}
 \end{center}
 \caption{The optimal mean photon number emitted by Alice ($\alpha_{A}$) that outputs Fig. \ref{F1}. The bold lines correspond to (a). See also the main text for the explanation. \label{F2}}
\end{figure}

We also plot the corresponding optimal $\alpha_A$ in Fig. \ref{F2}. Notice that the mean photon number increases in some regime in some cases of (a), and recall that this increase does not change $\Delta_{\rm ini}$. If we increased the intensity in scheme I with the distance, then we would have more basis-dependent flaw, resulting in shortening of the achievable distance. This may be an intuitive reason why we see no such increase in Figs. \ref{fig:intensity33}, \ref{fig:intensityimpferfectPMI}, and 9.

\begin{figure}
\begin{center}
 \includegraphics[scale=0.6]{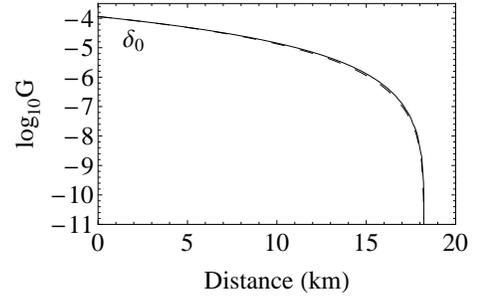}
 \end{center}
 \caption{The key generation rates of each setting as a function of the distance with $p_{\rm dark}=1.0\times10^{-7}$, $\eta_{{\rm det}, A}=\eta_{{\rm det}, B}=0.30$, and $\delta_0=0.063$. Note that we double $\eta_{{\rm det}, A}=\eta_{{\rm det}, B}$ compared to the one of \cite{uqcc}, or we effectively consider the polarization encoding \cite{LCQ11}. Dashed line: (a) MU is at Bob's side, i.e., $l_{B}=0$. Solid line: (b) MU is just in the middle between Alice and Bob. The key rates are almost superposed. See also the main text for the explanation. \label{F1new}}
\end{figure}

Like in the phase encoding scheme I, we investigate the feasibility of the phase encoding scheme II with the current technologies by replacing $p_{\rm dark}=8.5\times10^{-7}$, $\eta_{{\rm det}, A}=\eta_{{\rm det}, B}=0.045$, and $e_{\rm ali}=0.033$ with $p_{\rm dark}=1.0\times10^{-7}$, $\eta_{{\rm det}, A}=\eta_{{\rm det}, B}=0.15$ \cite{uqcc}, and $e_{\rm ali}=0.0075$ \cite{LCQ11}. With this upgrade, we have confirmed the impossibility of the key generation, however if we double the quantum efficiency of the detector or equivalently, if we assume the polarization encoding so that the factor of $1/2$, which is introduced by the phase-to-polarization converter, is removed both from $\eta_{A}$ and $\eta_{B}$ in Eq. (\ref{exp-data-II}), then we can generate the key, which is shown in Fig. \ref{F1new} (also see Fig. 13).

Finally, we note that our simulation is essentially the same as the polarization coding since the fact that we use phase encoding is only reflected by the dominator of 2 in $\eta_A$ and $\eta_B$ in Eq. (\ref{exp-data-II}). Thus, the behavior of the key generation rate against the degradation of the state preparation is the same also in polarization based MDIQKD. Also note that even in the standard BB84, $\Delta$ decays exponentially with increasing distance. Thus, we conclude that very precise state preparation is very crucial in the security of not only MDIQKD but also in standard QKD. We also note that our estimation of the fidelity might be too pessimistic since we have assumed that the degradation of the extinction ratio is only due to imperfect phase modulation. In reality, the imperfection of Mach-Zehnder interferometer and other imperfections should contribute to the degradation, and the fidelity should be closer to $1$ than the one based on our model.

\begin{figure}
\begin{center}
 \includegraphics[scale=0.57]{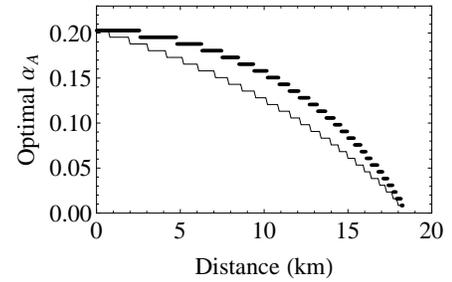}
 \end{center}
 \caption{The optimal mean photon number emitted by Alice ($\alpha_{A}$) that outputs Fig. \ref{F1new}. The bold lines correspond to (a). See also the main text for the explanation. \label{F2new}}
\end{figure}

\section{Summary and Discussion}\label{sec:summary}
In summary, we have proposed two phase encoding MDIQKD schemes. The first scheme is based on the phase locking technique and the other one is based on the conversion of the pulses in the standard phase encoding BB84 to polarization modes. We proved the security of the first scheme, which intrinsically possesses basis-dependent flaw, as well as the second scheme with the assumption of the basis-dependent flaw in the single-photon part of the pulses. Based on the security proof, we also evaluate the effect of imperfect state preparation, and especially we focus our attention to the imperfect phase modulation.

While the first scheme can cover relatively short distances of the key generation, this scheme has an advantage that the basis-dependent flaw can be controlled by the intensities of the pulses. Thanks to this property, we have confirmed based on a simple model that 3 or 5 times of the improvement in the accuracy of the phase modulation is enough to generate the key. Moreover, we have confirmed that the key generation is possible even without these improvements if we implement this scheme by using the up-to-date technologies and the control of intensities of the laser source is precise. On the other hand, it is not so clear to us how accurate we can lock the phase of two spatially separated laser sources, which is important for the performance of scheme I. Our result still implies that scheme I can tolerate up to some extent of the imperfect phase locking errors, which should be basically the same as the misalignment errors, but further analysis of the accuracy from the experimental viewpoint is necessary. We leave this problem for the future studies.

\begin{figure}
\begin{center}
 \includegraphics[scale=0.6]{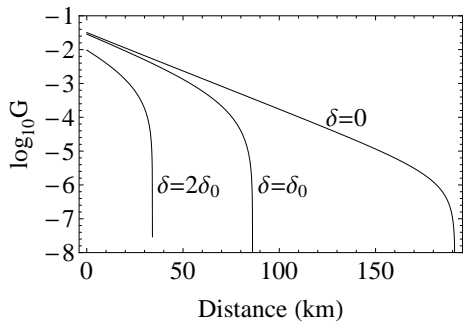}
 \end{center}
 \caption{The key generation rates of the standard BB84 with infinite decoy states from $X$-basis when $\delta=0$, $\delta=\delta_{0}$, $\delta=2\delta_{0}$ where $\delta$ is the amount of the phase modulation error. \label{sF1}}
\end{figure}

\begin{figure}
\begin{center}
 \includegraphics[scale=0.57]{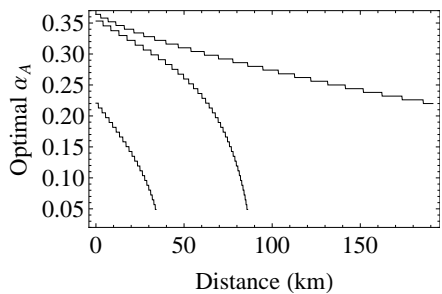}
 \end{center}
 \caption{The optimal mean photon number emitted by Alice ($\alpha_{A}$) that outputs Fig. \ref{sF1}. \label{sF2}}
\end{figure}

The second scheme can cover much longer distances when the fidelity of the {\it single-photon components} of $Y$-basis and $X$-basis density matrices is perfect or extremely close to perfect. When we consider the slight degradations of the fidelity, however, we found that the achievable distances drop significantly. This suggests that we need a photon source with a very high fidelity, and very accurate estimation of the fidelity of the single-photon subspace is also indispensable. 

In our estimation of the imperfection of the phase modulation, we simply assume that the degradation of the extinction ratio is only due to imperfect phase modulation, which might be too pessimistic, and the imperfection of Mach-Zehnder interferometer and other imperfections contribute to the degradation. Thus, the actual fidelity between the density matrices of the single-photon part in two bases might be very close to 1, which should be experimentally confirmed for the secure communication. We note that the use of the passive device to prepare the state \cite{passive} may be a promising way for the very accurate state preparation.

We remark that the accurate preparation of the state is very important not only in MDIQKD but also in standard QKD where Eve can enhance the imbalance of the quantum coin exponentially with the increase of the distance. To see this point, we respectively plot in Fig. \ref{sF1} and Fig. \ref{sF2} the key generation rate of standard BB84 with infinite decoy states in $X$-basis and its optimal mean photon number assuming $p_{\rm dark}=1.0\times10^{-7}$, $\eta_{{\rm det}, A}=0.15$ \cite{uqcc}, $e_{\rm ali}=0.0075$, $f(\delta_{x})=1.22$, and $\xi=0.21$. Again, $\delta_{0} \sim 0.063$ is the typical value of the phase modulation error, and we see in the figure that the degradation of the phase modulator in terms of the accuracy significantly decreases the achievable distance of secure key generation. One notices that standard decoy BB84 is more robust against the degradation since the probability that the measurement outcome of the quantum coin along $X$-basis is $\ket{1_x}$ given the successful detection of the signal by Bob is written as $\Delta_{\rm ini}=\frac{1}{2}(1-F_{A}^{(1)})$ rather than $\Delta_{\rm ini}=\frac{1}{2}(1-F_{A}^{(1)}F_{B}^{(1)})$. On the other hand, one has to remember that we trust the operation of Bob's detectors in this simulation, which may not hold in practice.

Finally, we neglect the effect of the fluctuation of the intensity and the center frequency of the laser light in our study, which we will analyze in the future works. In summary, our work highlights the importance of very accurate preparation of the states to avoid basis-dependent flaws.

\subsection{Acknowledgment}
We thank X. Ma, M. Curty, K. Azuma, T. Yamamoto, R. Namiki, T. Honjo, H. Takesue, Y. Tokunaga, and especially G. Kato for enlightening discussions. Part of this research was conducted when K. T and C-H. F. F visited the university of Toronto, and they express their sincere gratitude for all the supports and hospitalities that they received during their visit. This research is in part supported by the project ``Secure photonic network technology'' as part of ``The project UQCC'' by the National Institute of Information and Communications Technology (NICT) of Japan, in part by the Japan Society for the Promotion of Science (JSPS) through its Funding Program for World-Leading Innovative R$\&$D on Science and Technology (FIRST Program)", in part by RGC grant No. 700709P of the HKSAR Government, and also in part by NSERC, Canada Research Chair program, Canadian Institute for Advanced Research (CIFAR) and QuantumWorks.

\appendix

\section{Scheme I without noises and losses}
In this appendix, we give a detailed calculation about how scheme I works when there is no channel losses and noises. In order to calculate the joint probability that Alice and Bob obtain type-0 successful event, where only the detector D0 clicks, and they share the maximally entangled state $\ket{\Psi^{+}}$, we introduce a projector ${\hat \Pi}_{{\rm D_0}}\equiv{\hat P}\left(\ket{\overline{0}}_{{\rm D_0}}\right){\hat P}\left(\ket{0}_{{\rm D_1}}\right)$ that corresponds to type-0 successful event. Here, $\overline{0}$ represents the non-vacuum state. The state after Alice and Bob have the type-0 successful event ${\hat \Pi}_{{\rm D_0}}\ket{\zeta}_{A1, B1, {\rm D0}, {\rm D1}}$ (see Eq. (\ref{normal schI}) for the definition of $\ket{\zeta}_{A1, B1, {\rm D0}, {\rm D1}}$) can be expressed by 
\begin{eqnarray}
{\hat \openone_{A1, B1}}{\hat \Pi}_{D_0}\ket{\zeta}_{A1, B1, {\rm D0}, {\rm D1}}&=&\frac{a}{\sqrt{2}}\ket{\Phi^{+}}_{A1, B1}\ket{\phi_0}_{{\rm D0}}\ket{0}_{{\rm D1}}\nonumber\\
&+&\frac{b}{\sqrt{2}}\ket{\Psi^{+}}_{A1, B1}\ket{\phi_1}_{{\rm D0}}\ket{0}_{{\rm D1}}\,.\nonumber\\
\label{intermidiate}
\end{eqnarray}
Here, ${\hat \openone_{A1, B1}}$ is an identity operator on $A1$ and $B1$, $a$ and $b$ are complex numbers, and $\ket{\phi_0}$ and $\ket{\phi_1}$ are orthonormal bases, which are related with each other through
\begin{eqnarray}
{\hat P}\left(\ket{\overline{0}}\right)\ket{\sqrt{2\alpha'}}&\equiv&a\ket{\phi_0}+b\ket{\phi_1}\nonumber\\
{\hat P}\left(\ket{\overline{0}}\right)\ket{-\sqrt{2\alpha'}}&\equiv&a\ket{\phi_0}-b\ket{\phi_1}\,.
\end{eqnarray}
By a direct calculation, one can show that 
\begin{eqnarray}
|a|^2&=&\frac{(1-e^{-2\alpha'})^2}{2}\nonumber\\
|b|^2&=&\frac{(1-e^{-4\alpha'})}{2}\,.
\end{eqnarray}
Finally, by taking the partial trace over the system ${\rm D0}$ and ${\rm D1}$ in Eq. (\ref{intermidiate}), we can see that Alice and Bob share either $\ket{\Phi^{+}}$ or $\ket{\Psi^{+}}$ probabilistically, and 
the joint probability that they obtain type-0 successful event and share the maximally entangled state $\ket{\Psi^{+}}$ is given by $|b|^2/2$. In the same manner, we can readily calculate the other joint probabilities.

\section{Imperfection of the phase modulator}
In this appendix, we give an estimation of the fidelity between the density matrices in $X$ and $Y$ bases by using the extinction ratio. In this estimation, we assume that the source of the imperfections is only due to the imperfect phase modulation and the stability of the intensity control of the coherent light source is negligible. 

Imagine that we generate two pulses, one of which is the reference light and the other one of which is the signal light, and these pulses are spatially separated. Then, we input these pulses into a Mach-Zehnder interferometer, which is composed of two 50:50 beam splitters, and the output from one of the two output ports gives us the desired state and the other output port gives a wrong state. Let $T$ and $t$ be transmission rates of the pulses to the correct port and the wrong port, which satisfies 
\begin{eqnarray}
T+t=1\nonumber\\
\eta_{\rm ex}\equiv\frac{t}{T}\,,
\end{eqnarray}
where $\eta_{\rm ex}$ is the extinction ratio. In typical experiments, $\eta_{\rm ex}$ is in the order of $10^{-3}$ \cite{extinction}. We assume that the Mach-Zehnder interferometer is perfect and the imperfection of the extinction ratio is only due to imperfect phase modulations. 

Now, suppose that we plug $\ket{\sqrt{\alpha}}$ and $\ket{e^{i(\pi+\delta)}\sqrt{\alpha}}$ into the perfect Mach-Zehnder interferometer, where $\delta$ represents the imperfect the phase modulation when we want to apply phase shift of $\pi$. Since Mach-Zehnder interferometer is composed of 50:50 beam splitters, $\eta_{\rm ex}$ can be represented by $\left|\tan\frac{\delta}{2}\right|^2$, and thus we can obtain the imperfection of $\delta$ by solving the following equation 
\begin{eqnarray}
\left|\tan\frac{\delta}{2}\right|^2=\eta_{\rm ex}\,.
\label{imperfect phase modulator}
\end{eqnarray}
For instance, when $\eta_{\rm ex}=10^{-3}$, we have $|\delta|\sim0.063\equiv\delta_0$ that is equivalent to about $3.62$ degrees. We rely on this equation to estimate the accuracy of the phase modulator, and we assume that the actual phase modulation is $\theta+|\delta| \theta/\pi$, i.e., the degree of the imperfect phase modulation is proportional to the desired phase modulation. We remark that $3.62$ degrees seem rather large to us, and we believe that this can be substantially improved through careful calibration and/or engineering of the preparation process.

In the case of scheme I, the ideal density matrix for $X$ basis $\rho_{X}^{(\rm Ideal)}$ is $(\ket{\alpha}\bra{\alpha}+\ket{-\alpha}\bra{-\alpha})/2$ and the one for $Y$ basis $\rho_{Y}^{(\rm Ideal)}$ is $(\ket{i\alpha}\bra{i\alpha}+\ket{-i\alpha}\bra{-i\alpha})/2$. Based on our imperfect phase modulation model, we have the density matrix for the actually generated states in $X$ basis as
\begin{eqnarray}
\rho_{X}^{(\rm Act)}(\alpha, \delta)=\left(\ket{\sqrt{\alpha}}\bra{\sqrt{\alpha}}+\ket{-e^{i\delta}\sqrt{\alpha}}\bra{-e^{i\delta}\sqrt{\alpha}}\right)/2\nonumber\\
\end{eqnarray}
and the one for $Y$ basis as
\begin{eqnarray}
\rho_{Y}^{(\rm Act)}(\alpha, \delta)&=&\Big(\ket{i e^{i|\delta|/2}\sqrt{\alpha}}\bra{i e^{i|\delta|/2}\sqrt{\alpha}}\nonumber\\
&+&\ket{-i e^{-i|\delta|/2}\sqrt{\alpha}}\bra{-i e^{-i|\delta|/2}\sqrt{\alpha}}\Big)/2\,.\nonumber\\
\end{eqnarray}
Here, note that when we want to prepare $\ket{\sqrt{\alpha}}$, we do not apply any phase modulation.

In the case of scheme II, the ideal single-photon density matrix for $X$ basis $\rho_{X}^{({\rm Ideal},1)}$ is $(\ket{0_x}\bra{0_x}+\ket{1_x}\bra{1_x})/2$ and the one for $Y$ basis $\rho_{Y}^{({\rm Ideal},1)}$ is $(\ket{0_y}\bra{0_y}+\ket{1_y}\bra{1_y})/2$. With the assumption on the accuracy of the phase modulator, we have the density matrix for the actually generated states in $X$ basis as
\begin{eqnarray}
\rho_{X}^{({\rm Act}, 1)}&=&\Big({\hat P}\left[(\ket{0_z}+\ket{1_z})/\sqrt{2}\right]\nonumber\\
&+&{\hat P}\left[(\ket{0_z}-e^{i|\delta|}\ket{1_z})/\sqrt{2}\right]\Big)/2
\end{eqnarray}
and the one for $Y$ basis as
\begin{eqnarray}
\rho_{Y}^{({\rm Act}, 1)}&=&\Big({\hat P}\left[(\ket{0_z}+i e^{i|\delta|/2}\ket{1_z})/\sqrt{2}\right]\nonumber\\
&+&{\hat P}\left[(\ket{0_z}-i e^{-i|\delta|/2}\ket{1_z})/\sqrt{2}\right]\Big)/2\,.
\end{eqnarray}

\section{Erratum: Phase encoding schemes for measurement device independent quantum key distribution and basis-dependent flaw [Phys. Rev. A 85, 042307 and arXiv:1111.3413]}

We would like to correct Eq. (9) in our paper and, subsequently, modify figures for the key generation rates. 
These corrections do not affect the validity of the main conclusions reported in the paper. 
The correct form of Eq. (9) in our paper [K. Tamaki, H-K. Lo, C-H. F. Fung, and B. Qi, Phys. Rev. A 85, 042307 (2012) and arXiv:1111.3413] 
should be 
\begin{eqnarray}
\Delta_{\rm ini}\equiv {\rm Min}\left(\Delta_{\rm ini}^{(A)}, \Delta_{\rm ini}^{(B)}\right)\,,
\label{0}
\end{eqnarray}
where
\begin{eqnarray}
\Delta_{\rm ini}^{(A)}&\equiv& \Big[1-{\rm Max}_{\theta, \xi_{Y}^{(A)}, \xi_{X}^{(A)}}{\rm Re}\Big(e^{\mathrm{i}\theta}
\Big\langle\Psi_{Y,\xi_{Y}^{(A)}}^{(A)} \Big|\Psi_{X,\xi_{X}^{(A)}}^{(A)}\Big\rangle\Big)\nonumber\\
&\times&F\left(\rho_{B}^{(X)}, \rho_{B}^{(Y)}\right)\Big]/2\nonumber\\
\label{1}
\end{eqnarray}
or 
\begin{eqnarray}
\Delta_{\rm ini}^{(B)}&\equiv& \Big[1-{\rm Max}_{\theta, \xi_{Y}^{(B)}, \xi_{X}^{(B)}}{\rm Re}\left(e^{\mathrm{i}\theta}\Big\langle\Psi_{Y,\xi_{Y}^{(B)}}^{(B)} \Big|\Psi_{X,\xi_{X}^{(B)}}^{(B)}\Big\rangle\right)\nonumber\\
&\times&F\left(\rho_{A}^{(X)}, \rho_{A}^{(Y)}\right)\Big]/2\,.\nonumber\\
\label{2}
\end{eqnarray}
Here, $0\le\theta<2\pi$, $0\le\xi_{W}^{(A)}<2\pi$, and $0\le\xi_{W}^{(B)}<2\pi$, and $\ket{\Psi_{W,\xi_{W}^{(A)}}^{(A)}}$ is defined by
\begin{eqnarray}
\ket{\Psi_{W,\xi_{W}^{(A)}}^{(A)}}=\frac{1}{\sqrt{2}}\left(\ket{0_{W}}_{Aq}\ket{\chi_{0W}^{(A)}}+e^{\mathrm{i}\xi_{W}^{(A)}}\ket{1_{W}}_{Aq}\ket{\chi_{1W}^{(A)}}\right)\,,\nonumber\\
\label{3}
\end{eqnarray}
where 
$\ket{\chi_{iW}^{(A)}}$ is a purification of $\rho^{(A)}_{iW}$, which is the state that Alice actually prepares for the bit 
value $i$ in basis $W(={\rm X}, {\rm Y})$, and $Aq$ is Alice's qubit system. One can choose any purification for $\ket{\chi_{iW}^{(A)}}$, 
and in particular it should be chosen in such a way that it maximizes the inner product in Eq. (C2) or (C3). One can similarly define 
$\ket{\Psi_{W,\xi_{W}^{(B)}}^{(B)}}$, and $\theta$ is introduced via considering a joint state 
involving the quantum coin as
\begin{eqnarray}
\ket{\Psi_{\theta, \xi^{(A)}, \xi^{(B)}}}&\equiv&\frac{1}{\sqrt{2}}\Big(\ket{0_z}_{C}\ket{\Psi_{X,\xi_{X}^{(A)}}^{(A)}}\ket{\Psi_{X,\xi_{X}^{(B)}}^{(B)}}\nonumber\\
&+&e^{\mathrm{i}\theta}\ket{1_z}_{C}\ket{\Psi_{Y,\xi_{Y}^{(A)}}^{(A)}}\ket{\Psi_{Y,\xi_{Y}^{(B)}}^{(B)}}\Big)\,.
\label{coin-joint-state-post-selected}
\end{eqnarray}
Due to this change, the figures for the key generation rate have to be revised. As the examples of revised figures, 
we show the revised version of Figs. 8, 9, 12, and 13, which are the most important figures for our main conclusions to hold. 
Notice that there are only minor changes in Figs. 8 and 9 and the changes in Figs. 12 and 13 are relatively big. However, 
the big changes do not affect the validity of the main conclusions in our paper, which is the importance of the state
preparation in MDIQKD and the fact that our schemes can generate the key with the practical channel mode that we have assumed.

\begin{figure}
\begin{center}
 \includegraphics[scale=0.6]{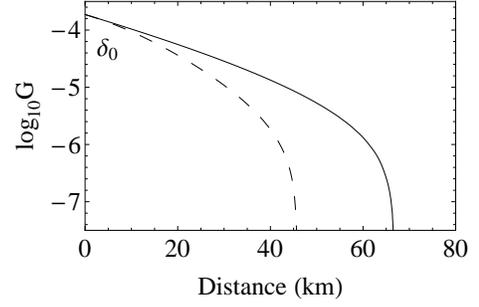}
 \end{center}
 \caption{The revised figure for Fig. 8 in our paper.
\label{fig:keyimpferfectPMInew}}
\end{figure}

\begin{figure}
\begin{center}
 \includegraphics[scale=0.6]{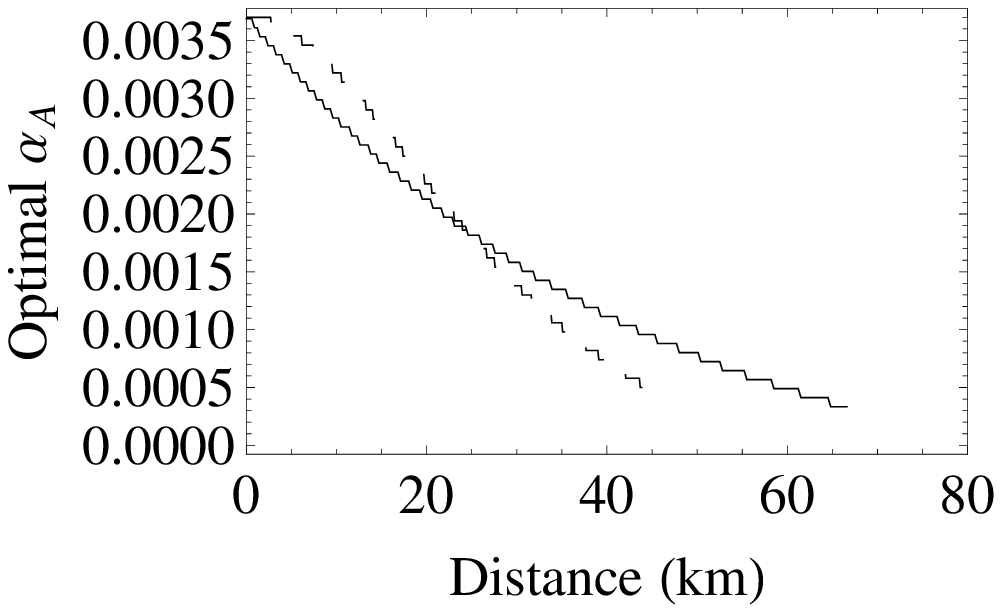}
 \end{center}
 \caption{The revised figure for Fig. 9 in our paper. \label{fig:intensityimpferfectPMInew}}
\end{figure}

\begin{figure}
\begin{center}
 \includegraphics[scale=0.6]{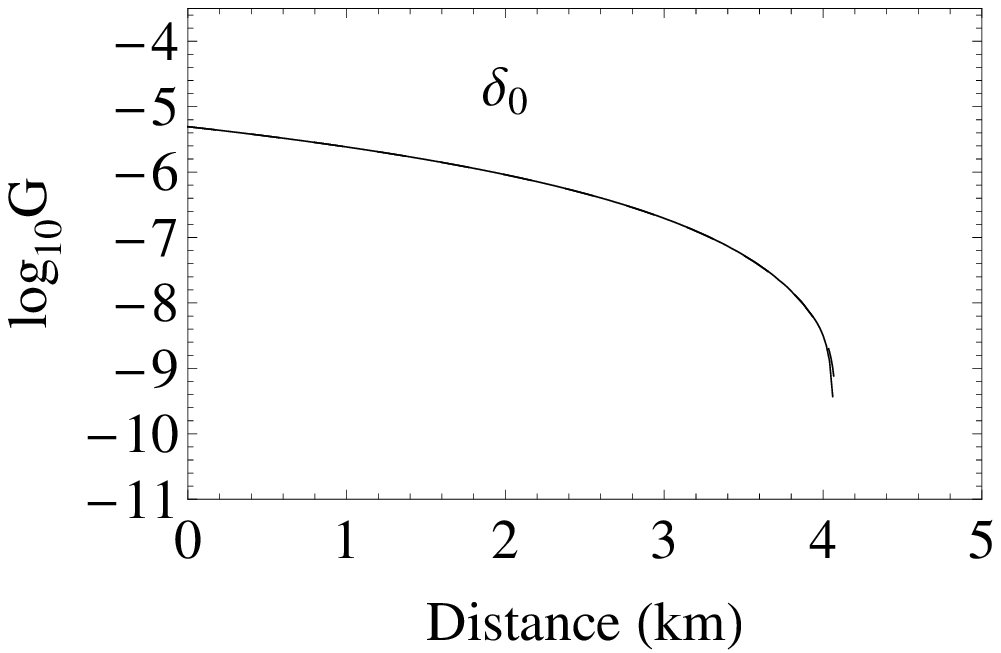}
 \end{center}
 \caption{The revised figure for Fig. 12 in our paper. Two lines are superposed.
\label{fig:keyimpferfectPMInew}}
\end{figure}

\begin{figure}
\begin{center}
 \includegraphics[scale=0.6]{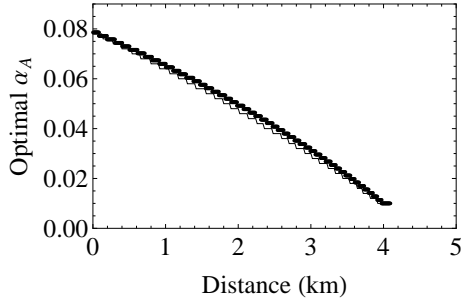}
 \end{center}
 \caption{The revised figure for Fig. 13 in our paper. Two lines are superposed. \label{fig:intensityimpferfectPMInew}}
\end{figure}

{\it Justification of Eq.~(C1)}

For the derivation of Eq.~(\ref{0}), we invoke Koashi's proof \cite{Koashi}. 
To apply Koashi's proof, it is important to ensure that
i) one of the two parties holds a virtual *qubit* (rather than a higher 
dimensional system) and ii) the fictitious measurements performed on the 
virtual qubit have to form *conjugate* observables. Therefore, it is not 
valid to consider fidelity alone (which allows arbitrary purifications that 
may not satisfy the conjugate observables requirement). Fortunately, it turns out 
to be easy to modify our equation to satisfy the above two requirements.

Since the difference between Eq.~(C2) 
and Eq.~(C3) comes from whether we consider Alice's virtual qubit or Bob's virtual qubit, we
focus only on Eq.~(C2) and the same argument holds for Eq.~(C3). In Koashi's proof, the security is 
guaranteed via two alternative tasks, (i) agreement on X (key distillation basis) and (ii)
Alice's or Bob's preparation of an eigenstate of Y, the conjugate basis of X, 
with use of an extra communication channel. The problem with the original (i.e. uncorrected) 
version of Eq.~(9) is the following. If we use the uncorrected version of Eq.~(9) in our paper, 
then the use of the fidelity means that the real part in Eq.~(C2)
is equivalent to $\left|\Big\langle\Psi_{Y,0}^{(A)}\Big| \left(U_{Aq}\otimes\openone\Big|\Psi_{X,0}^{(A)}\Big\rangle\right)\right|$
with the maximization over {\it all possible} local unitary operators $U_{Aq}$. %Let $U_{Aq}^{*}$ be the optimal unitary operator.
In this case, if Alice performs a measurement along X basis, then it violates the %Alice has to measure along $U_{Aq}^{*}X U_{Aq}^{*\dagger}$ to preserve the 
correspondence between
her sending state $\rho^{(A)}_{iW}$ and her qubit state $\ket{i_{W}}_{Aq}$ in general, and thus, in the uncorrected 
version of Eq. (9) in our paper, the argument
based on the fidelity does not guarantee the security of the protocol. %Since $U_{Aq}^{^*}X U_{Aq}^{*\dagger}$
%is not a conjugate basis of $Y$ in general, we cannot apply Koashi's proof. 
In contrast, with the corrected version of Eq.~(9) in our paper, since the maximization over $\theta$ and 
$\xi_{W}^{(A)}$ in Eq.~(C2) preserves the relationship between Alice's sending state and her qubit state as well as the conjugate relationship between X and Y, 
we can apply Koashi's proof for the security argument of the protocol.

We thank X. B. Wang \cite{Wang} for raising the concern about the validity of Eq.~(9) in our paper.


\begin{thebibliography}{}
\bibitem{Bell state} The definition of the four Bell state is as follows. $\ket{\Psi^{+}}\equiv\frac{1}{\sqrt{2}}[\ket{0_z}_{A1}\ket{1_z}_{B1}+\ket{1_z}_{A1}\ket{0_z}_{B1}]=\frac{1}{\sqrt{2}}[\ket{0_x}_{A1}\ket{0_x}_{B1}-\ket{1_x}_{A1}\ket{1_x}_{B1}]$, $\ket{\Phi^{-}}\equiv\frac{1}{\sqrt{2}}[\ket{0_z}_{A1}\ket{0_z}_{B1}-\ket{1_z}_{A1}\ket{1_z}_{B1}]=\frac{1}{\sqrt{2}}[\ket{0_x}_{A1}\ket{1_x}_{B1}+\ket{1_x}_{A1}\ket{0_x}_{B1}]$, $\ket{\Phi^{+}}\equiv\frac{1}{\sqrt{2}}[\ket{0_z}_{A1}\ket{0_z}_{B1}+\ket{1_z}_{A1}\ket{1_z}_{B1}]=\frac{1}{\sqrt{2}}[\ket{0_x}_{A1}\ket{0_x}_{B1}+\ket{1_x}_{A1}\ket{1_x}_{B1}]$, and $\ket{\Psi^{-}}\equiv\frac{1}{\sqrt{2}}(\ket{0_z}_{A1}\ket{1_z}_{B1}-\ket{1_z}_{A1}\ket{0_z}_{B1})=\frac{1}{\sqrt{2}}(\ket{0_x}_{A1}\ket{1_x}_{B1}-\ket{1_x}_{A1}\ket{0_x}_{B1})$.
\bibitem{Y basis} In the case of $Y$-basis, we replace the following discussion with the one starting with $\ket{\phi_{y}^{(+)} (\ket{-i\sqrt{\alpha_{A}}})}\ket{\phi_{y}^{(+)} (\ket{-i\sqrt{\alpha_{B}}})}$ instead of $\ket{\phi_{x}^{(+)} (\ket{\sqrt{\alpha_{A}}})}\ket{\phi_{x}^{(+)} (\ket{\sqrt{\alpha_{B}}})}$. See Eq. (\ref{def-state}) for the definition of the state.
\bibitem{intuition} One of the most simplest proofs is Shor-Preskill's proof \cite{SP00}. The intuition of this proof is as follows. Note that if Alice and Bob share some pairs of $\ket{\Psi^{+}}$, (i.e., Alice has one half of each pair and Bob has the other half), then they can generate a secure key by performing $X$-basis measurement. The reason of the security is that this state is a pure state, which means that this state has no correlations with the third system including Eve's system. Due to the intervention by Eve, Alice and Bob do not share this pure state in general, but instead they share noisy pairs. The basic idea of the proof is to consider the distillation of $\ket{\Psi^{+}}$ from the noisy pairs. 

\,\,For the distillation, note that $\ket{\Psi^{+}}$ is only one qubit pair state that has no bit errors in $X$-basis (we call this error as the bit error) and has no bit errors in $Y$-basis (we call this error as the phase error). It is known that if Alice and Bob employ the so-called CSS code (Calderbank-Shor-Steane code) \cite{CSS}, then the noisy pairs are projected to a classical mixture of the four Bell states, i.e., $\ket{\Psi^{+}}$, $\ket{\Phi^{+}}$ ($\ket{\Psi^{+}}$ with the phase error), $\ket{\Psi^{-}}$ ($\ket{\Psi^{+}}$ with the bit error), and $\ket{\Phi^{-}}$ ($\ket{\Psi^{+}}$ with both the phase and bit errors). Moreover, if Alice and Bob choose a correct CSS code, which can be achieved by random sampling procedure, then CSS code can detect the position of the erroneous pair with high probability. Thus, by performing bit and phase flip operation depending on the detected error positions, Alice and Bob can distill some qubit pairs that are very close in fidelity to the product state of $\ket{\Psi^{+}}$. In general, implementation of the above scheme requires a quantum computer. Fortunately, Shor-Preskill showed that the bit error detection and bit flip operation can be done classically, and the phase error detection and phase flip operation need not be done, but exactly the same key can be obtained by  the privacy amplification, so that we do not need to possess a quantum computer for the key distillation.
\bibitem{Y basis2} We have chosen $\ket{\phi_{y}^{(+)} (\ket{-i\sqrt{\alpha_{A}}})}\ket{\phi_{y}^{(+)} (\ket{-i\sqrt{\alpha_{B}}})}$ rather than $\ket{\phi_{y}^{(+)} (i\sqrt{\alpha_{A}})}\ket{\phi_{y}^{(+)} (i\sqrt{\alpha_{B}})}$ to maximize the inner product with $\ket{\phi_{x}^{(+)} (\ket{\sqrt{\alpha_{A}}})}\ket{\phi_{x}^{(+)} (\ket{\sqrt{\alpha_{B}}})}$. This large inner product is important for the security proof.
\end{thebibliography}
\end{document}